\documentclass[]{spie}  

\usepackage{amsmath,amsfonts,amssymb}
\usepackage{graphicx}
\usepackage[colorlinks=true, allcolors=blue]{hyperref}
\usepackage{xcolor}
\usepackage{aas_macros}
\usepackage[normalem]{ulem}

\newcommand{\n}{\hat{\bf n}}

\usepackage[normalem]{ulem} 

\def\bmode{B-mode}
\def\TtoQU{$T\rightarrow Q,U$}
\def\TtoP{$T\rightarrow P$}

\title{The Canadian Galactic Emission Mapper: A New 8-10\,GHz Telescope to Map the Polarization of the Northern Sky}

\author[a,b]{Joshua MacEachern}
\author[b]{Mandana Amiri}
\author[c]{Charles L. Bennett}
\author[b]{Guinevere Berg}
\author[b]{Mark Halpern}
\author[b]{Gary Hinshaw}
\author[a]{Gordon Lacy}
\author[b]{Thomas J. Rennie}
\author[b]{Shuyu van Kerkwijk}
\author[a]{Bruce Veidt}
\author[b]{Pedro Villalba-González}
\author[c]{Janet Weiland}
\author[b]{Don Wiebe}
\author[d]{Edward J. Wollack}
\author[b]{Parham Zarei}

\affil[a]{Dominion Radio Astrophysical Observatory, Herzberg Astronomy \& Astrophysics Research Centre, National Research Council Canada,
Kaleden, BC V0H 1K0, Canada}
\affil[b]{Department of Physics and Astronomy, University of British Columbia, Vancouver, BC V6T 1Z1, Canada}
\affil[c]{The William H. Miller III Department of Physics and Astronomy,
Johns Hopkins University, Baltimore, MD 21218, USA}
\affil[d]{NASA Goddard Space Flight Center, Greenbelt, MD 20771, USA}
\authorinfo{Further author information: (Send correspondence to P.Z or P.V.G.)\\P.Z: E-mail: pzarei@phas.ubc.ca\\  P.V.G.: E-mail: pedrovg@phas.ubc.ca}

\authorinfo{Further author information: (Send correspondence to J. MacEachern)\\J. MacEachern: E-mail: joshua.maceachern@nrc-cnrc.gc.ca}

\begin{document} 
\maketitle
\begin{abstract}
The Canadian Galactic Emission Mapper (CGEM) radio telescope is mapping polarized Galactic foregrounds from 8-10\,GHz to aid in the search for B-modes in the Cosmic Microwave Background (CMB). Here we describe the design and early on-sky characterization of the CGEM optics. CGEM employs an on-axis, hat feed optical design that is a body of revolution (BOR) and exhibits excellent intrinsic polarization purity. We describe how we have optimized the optics with a novel framework that can directly minimize simulated intensity to polarized intensity ($T\rightarrow P$) leakage in angular power spectrum space. The optimized optics exhibit simulated $T\rightarrow P$ leakage that is orders of magnitude below anticipated $C_\ell^{BB}$ for $r = 10^{-3}$ when scaled from 8-10\,GHz to the CMB observing window near 95\,GHz. We go on to describe the mechanical design of the optics, including a novel secondary mirror support made from Astroquartz composite that has low loss, low dielectric constant, and preserves the BOR symmetry of the optics. We then showcase the early on-sky performance of the optics with observations of the Sun and satellites, which probe the beam to 40\,dB down from the peak. A beam model for the deployed CGEM optics based on electromagnetic simulations is in excellent agreement with the on-sky data.

\end{abstract}


\section{Introduction}\label{sec:intro}

Cosmic inflation is an essential part of the standard model of cosmology. This early period of accelerated expansion could solve the horizon, flatness, and magnetic monopole problems, and would provide a natural means of production for the Gaussian, adiabatic, scalar metric perturbations that source the well-measured temperature and E-mode polarization anisotropies in the Cosmic Microwave Background (CMB) \cite{particle_data_group_review}. Many models of inflation predict that it would also produce a nearly scale-invariant distribution of tensor metric perturbations (gravitational waves), and these could have left a detectable signature in the parity-odd, B-mode component of the polarization of the CMB. As B-modes in the CMB are uniquely sourced by tensor metric perturbations from inflation, their detection would be strong evidence that inflation occurred and would provide a measurement of the energy scale of inflation ($\sim$$10^{15}$-$10^{16}$\,GeV) \cite{particle_data_group_review, quest_for_b_modes}. This would probe physics at energy scales that are otherwise inaccessible at present.

Numerous experiments are searching for B-modes in the CMB and have placed stringent upper limits. BICEP is the most sensitive, with an upper limit on the primordial tensor-to-scalar ratio of $r < 0.036$ \cite{BICEP} (corresponding to polarization fluctuations at sub-nK$_\text{cmb}^2$ levels). These instruments must accordingly have high sensitivity and must map polarization with minimal systematics on large-angular scales (degree scales or larger), where foreground lensing of CMB E-modes into B-modes is less important. Another crucial challenge is the presence of polarized emission from our Galaxy that must be mapped in detail and at as many frequencies as possible to facilitate foreground separation \cite{quest_for_b_modes, planck_diffuse, clive_foreground_review, BICEP, remazailles_dickinson_foregrounds}. B-modes, if present, are already constrained to be dominated by polarized foreground emission at all frequencies \cite{planck_diffuse}, highlighting the importance of accurate polarized foreground maps. Diffuse polarized foreground components are thermal dust (brighter at higher frequencies), synchrotron emission (brighter at lower frequencies), and possibly spinning dust emission if it is polarized (peaking near $\sim$20-30\,GHz in flux-density units)\cite{planck_diffuse, clive_foreground_review, remazailles_dickinson_foregrounds}. As sensitivity levels increase and upper limits on $r$ are pushed lower and lower (with the next generation of CMB experiments targeting upper limits of $r < 0.001$\cite{cmb_s4_science_book}), mapping polarized foregrounds will only become more important. Polarized synchrotron, which is still poorly understood as a CMB foreground, will be crucial to measure \cite{weiland}. This emission is produced by relativistic cosmic ray electrons spiraling around Galactic magnetic field lines. Its frequency scaling can vary spatially due to variations in the energy distribution of cosmic ray electrons, variations in the Galactic magnetic field strength and orientation, and synchrotron self absorption. This complexity necessitates mapping synchrotron at as many frequencies as possible in order to make CMB polarization measurements robust to contamination from synchrotron B-modes\cite{weiland}.

We have built the Canadian Galactic Emission Mapper (CGEM) to map the Northern sky from 8-10\,GHz to improve models of low-frequency, polarized Galactic foregrounds to aid in the B-mode search. CGEM is a 4\,m, single-dish, radio telescope situated at the Dominion Radio Astrophysical Observatory near Penticton, British Columbia, Canada. The site is in a protected, radio-quiet valley that is an excellent site for such observations. CGEM's 4\,m primary mirror gives it a modest angular resolution of 0.5$^\circ$, which is sufficient for measuring large scales for the \bmode\ search. The instrument maps the sky by continuously scanning at 2\,rpm in azimuth and at fixed elevation. The high scan speed means that the boresight sweeps out large angular scales on the sky on short timescales, over which atmospheric and instrumental properties are stable. This allows large scales on the sky to be recovered more accurately during map making. An important benefit of this scan geometry is that every pixel is observed at 2 parallactic angles that are $\sim$$90^\circ$ apart. This modulates measured $Q$ and $U$ in the data and hence allows for more accurate polarization maps. CGEM sits inside a reflective ground shield that highly attenuates bright ground emission, which prevents it from introducing significant polarization systematics into the data and from substantially increasing the system noise level.

By observing near 9\,GHz, CGEM will isolate and map polarized synchrotron emission in a frequency band which is low enough that synchrotron is bright, but high enough that the data suffer negligible Faraday polarization rotation that occurs at lower frequencies. Observations in this band also have the power to constrain whether or not spinning dust emission is polarized, and can help separate the morphology of free-free emission, spinning dust emission, and synchrotron emission in intensity. At 8-10\,GHz we also avoid Starlink satellite uplink/downlink bands from 10.7-12.7\,GHz, and largely avoid an atmospheric liquid water emission/absorption line at 22\,GHz.

We have purpose-built CGEM to measure large-scale polarization at high sensitivity and with minimal systematics:
\begin{itemize}
    \item The optics were chosen and optimized to exhibit excellent polarization purity (the subject of this paper).
    \item The waveguide network, including an orthomode transducer (OMT), has been designed to give the cleanest possible separation of orthogonal polarizations.
    \item The low-noise amplifiers (LNAs) will be cryogenically cooled to give the system $T_{\text{sys}} \sim 15$\,K (including all losses, the atmosphere, and CMB).
    \item The superheterodyne receiver has phase-matched components, several stages of filtering, and will be temperature controlled.
    \item The correlator divides the 8-10\,GHz frequency band into 2048 frequency channels, each of which is sampled at up to 0.4\,ms cadence. This allows for high-cadence, narrow-band, radio-frequency interference (RFI) to be surgically removed from the data with minimal data loss. 
    \item The system has been designed to detect and correlate circular polarization, which results in cleaner measurements of Stokes $Q$ and $U$ that are unaffected by noise bias after correlation.
\end{itemize}

We have been mapping the sky with a non-cryogenic, ``warm radiometer" version of CGEM since February 2025. We are using this system as a testbed to understand our most important systematics so that we can improve the design for the cryogenic system. Here we describe the CGEM optical design and optical characterization effort so far. We have designed and optimized the CGEM warm radiometer optics to maximize polarization purity with a novel optimization framework. The mechanical design of the optics was carefully carried out to realize this simulated, optimized performance. Initial commissioning data suggests that the optics are performing very close to their design specification. There are small optical misalignments, but these are measured, correctable, and are shown to minimally affect the polarization purity of the optics.

This paper is organized as follows. In Section~\ref{sec:optical_design_and_optimization}, we describe the hat feed optical design and a framework for optimizing the design to minimize systematics. We then showcase the optimized performance. In Section~\ref{sec:mech}, we describe the mechanical design of the optics and how each optical component was manufactured. In Section~\ref{sec:on_sky_performance}, we describe several small, measured, and correctable optical misalignments and show simulations that suggest these do not significantly affect the polarization purity of the optics. We go on to showcase beam measurements using the Sun and Earth-observation satellites, and highlight the striking agreement between these measurements and a beam model based off of electromagnetic simulations. We then give concluding remarks in Section~\ref{sec:conclusions}.

\section{Optical Design and Optimization}\label{sec:optical_design_and_optimization}

\subsection{The Hat Feed Optical Design }\label{sec:hat_feed_intro}

\begin{figure} [ht]
   \begin{center}
   \begin{tabular}{c} 
   \includegraphics[height=5cm]{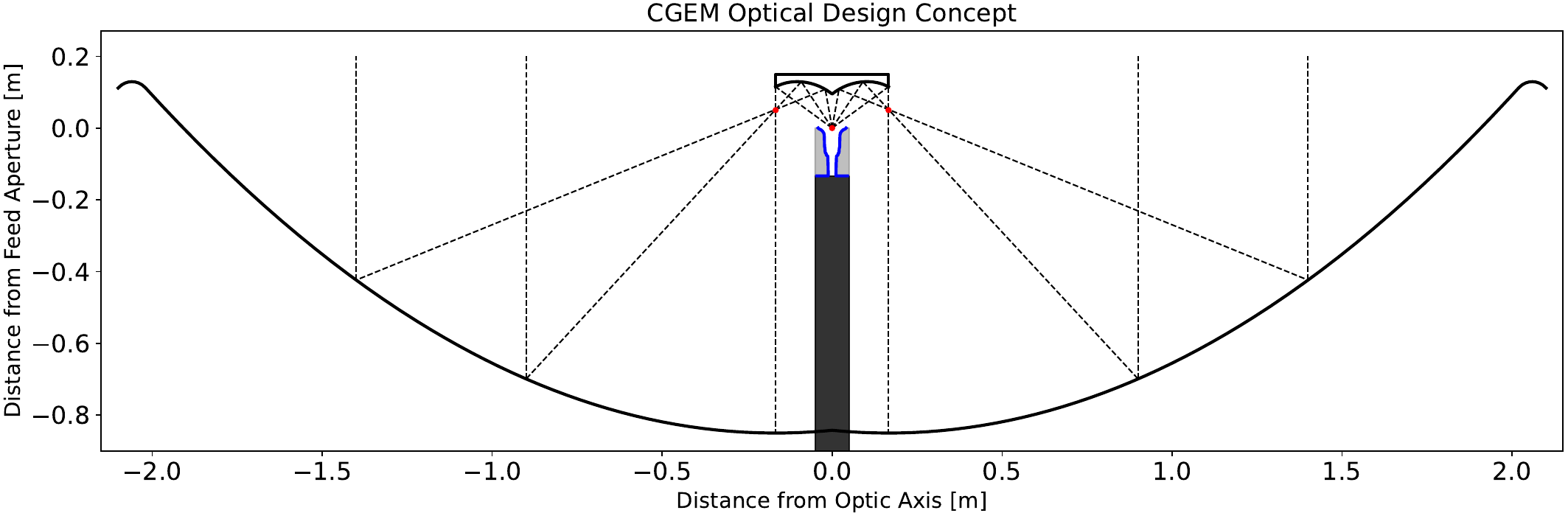}
   \end{tabular}
   \end{center}
   \caption[example] 
   { \label{fig:hat_feed_schem} 
A cross-section of a hat feed optical design. The y-axis shows the distance measured from the feed aperture and the x-axis shows the distance from the optic axis. The solid lines represent reflecting surfaces. The grey shaded region denotes the feed column which contains a feed horn (shown explicitly) and the rest of our waveguide network (not shown). The dashed lines show typical ray paths. In a geometric optics treatment, rays are focused to a ring focus (outer red dots) by the primary mirror and are refocused by the secondary mirror to the centre of the horn aperture (central red dot).}
\end{figure}

We have employed a hat feed optical design\cite{cutler, kildal, parabolic_hat_feed} for CGEM. A cross-section of a hat feed system is shown schematically in Figure \ref{fig:hat_feed_schem}. This is a dual-reflector, Gregorian optical system. In the receiving mode, a primary mirror collects signals from the sky and directs them onto a secondary mirror, which then re-directs them into a feed horn that carries the signals to the rest of the waveguide network and receiver. The primary mirror shape is obtained by offsetting a parabola radially from the optic axis and revolving it about the optic axis. The primary mirror thus has a ring focus, which is denoted by the two offset red dots in Figure \ref{fig:hat_feed_schem}. The secondary mirror shape is obtained by revolving a section of an ellipse about the optic axis. The geometry of the ellipse is chosen so that one focus overlaps with the focus of the primary and the other focus overlaps with the centre of the feed aperture (central red dot in Figure \ref{fig:hat_feed_schem}). In a geometric optics treatment, this means that an on-axis plane wave from the sky is focused to a ring by the primary mirror and refocused by the elliptical secondary to the centre of the feed aperture. In this way, the path length from the sky to the feed is the same for all rays entering the primary mirror aperture parallel to the optic axis. We use a smooth-walled feedhorn (described in Section \ref{sec:opt_goals_and_hat_feed_params}) that couples sky signals into a circular waveguide operating in the fundamental $TE_{11}$ mode from 8-10\,GHz. The secondary mirror is supported by a radio-transparent material (not shown in Figure \ref{fig:hat_feed_schem}) that is attached to the outside of the central feed column (the black region in Figure \ref{fig:hat_feed_schem}). The support and feed column are discussed in Section~\ref{sec:hat_support} and~\ref{sec:feed_column}.

In addition to having a clean beam, the ring-focus nature of the system also means the inner region of the secondary mirror (which is the region most strongly illuminated by the feedhorn) illuminates the outer region of the primary, and the outer region of the secondary illuminates the inner region of the primary. This means that most of the power radiated by the horn is spread out onto the primary mirror rather than back into the horn, meaning that these systems have low $S_{11}$. The on-axis nature of the hat feed also makes the optics compact and lightweight compared to other offset-reflector, polarization-pure designs with minimal aperture blockage (e.g. crossed-Dragone systems). The symmetry of the hat feed makes it very fast to simulate with numerical electromagnetic simulation software: it takes less than a second to determine $S_{11}$ and the far-field radiation patterns at one frequency for a CGEM-like instrument with TICRA Tools \cite{ticra}. The lightweight optics also mean that it is possible to scan at high speed without requiring an expensive mount/pointing system. One notable limitation of the hat feed is that it is a single-pixel system: additional detectors/receivers cannot be added at the focus to increase sensitivity. Accordingly, hat feed systems are well-suited to measuring relatively bright signals, such as polarized Galactic synchrotron emission in the case of CGEM.

The hat feed is a novel optical design in cosmology and radio astronomy, but hat feeds have been used in other areas, such as telecommunications. It is common to perturb the shapes of the reflecting surfaces in hat feed systems to optimize performance (see Ref.~\citenum{ticra_shaped_hat_feed}, for example). This was done extensively for the CGEM optics, as we shall see in the following subsections.

\subsection{Optimization Framework}\label{sec:opt_framework}
We have optimized the reflector and feedhorn shapes of the CGEM hat feed optics to further enhance the polarization purity. We use a novel optimization framework that directly minimizes simulations of our most important beam-related systematic: intensity-to-polarization leakage. We describe this framework in the following subsections. Note that this framework could be applied in the optical design of other instruments to minimize other beam-related systematics.

\subsubsection{Modelling CGEM Correlation Products and $T\rightarrow P$ Leakage}
The optimization framework heavily depends on a model for the CGEM correlation products, which we develop here. We base the formalism off of Ref.~\citenum{shaw}, but adapt it to the case of a single-dish radio telescope.

CGEM detects and correlates right- and left-hand circular polarization. In the circular basis, Stokes $Q$ and $U$ on the sky end up in the cross-correlation products, which are unaffected by noise bias and hence give cleaner measurements of linear polarization. We write the radiation patterns for the antenna port that predominantly measures circular polarization $p \in [r,l]$ as $\textbf{A}_p(\nu, \n) = A_p^{r}(\nu, \n)\hat{\textbf{r}} + A_p^{l}(\nu, \n)\hat{\textbf{l}}$, where, going forward, we use $r$ and $l$ in superscripts of radiation patterns to denote polarization vector components and $r$ and $l$ in subscripts to denote port labels. The signal measured by an antenna at a single frequency is an average of the electric field density, weighted by the antenna radiation pattern
\begin{align}
	W_p(\nu, t) &= \frac{1}{\sqrt{\Omega_p}} \int d^2\n \ \textbf{A}_p(\nu, \n, t) \cdot \boldsymbol{\epsilon}(\nu, \n),\label{eq:antenna_voltage}
\end{align}
where $\Omega_p = \int d^2\n \ |\textbf{A}_p(\n)|^2$ is a normalization factor and $\boldsymbol{\epsilon}(\nu, \n)$ is the electric field density, which is the electric field at the location of the antenna per unit frequency and per unit solid angle. We have added a time dependence to the radiation patterns to represent the telescope scanning the sky. The Stokes parameters on the sky can be written in terms of the electric field density as
\begin{align}
\begin{split}
	T &= \langle |\epsilon_r|^2 \rangle + \langle |\epsilon_l|^2 \rangle\\
	Q &= 2\,\text{Re}\left(\langle \epsilon_r \epsilon_l^*\rangle\right)\\
	U &= -\, 2 \, \text{Im}\left(\langle \epsilon_r \epsilon_l^* \rangle\right)\\
	V &= \langle |\epsilon_r|^2 \rangle - \langle |\epsilon_l|^2 \rangle\label{eq:stokes},
\end{split}
\end{align}
where we use $T$ to symbolize Stokes $I$ to make it more explicit that we are working in brightness temperature units.
Considering only antenna effects, the correlation products that are measured by the telescope then take the form
\begin{align}
	\langle W_{p}(\nu, t) W_{q}^*(\nu, t) \rangle \equiv W_{pq}(\nu, t) &= \frac{1}{\Omega_{pq}} \int d^2\n \ A^a_p(\nu, \n, t)A^{b*}_q(\nu, \n, t)\ \langle\epsilon_a(\nu, \n)\epsilon^*_b(\nu, \n)\rangle\label{eq:W_before_stokes},
\end{align}
where we have assumed that the electric field density is incoherent in position and frequency, and we are using Einstein summation notation to sum over the polarization vector component indices $a$ and $b$. We've also used the notation $\Omega_{pq} \equiv \sqrt{\Omega_p\Omega_q}$. From Equation \ref{eq:stokes}, this can be written in terms of the Stokes parameters as
\begin{align}
	W_{pq}(\nu, t) &= \int d^2\n\big[B_{pq}^{T}(\nu, \n, t) T(\nu, \n) + B_{pq}^{Q}(\nu, \n, t)  Q(\nu, \n) + B_{pq}^{U}(\nu, \n, t) U(\nu, \n) + B_{pq}^{V}(\nu, \n, t)  V(\nu, \n)\big],\label{eq:W_pq}
\end{align}
where
\begin{align}
\begin{split}
	B_{pq}^{T} &= \frac{1}{2\,\Omega_{pq}}\left( A_p^r A_q^{r*} + A_p^l A_q^{l*} \right) (\nu, \n, t)\\
	B_{pq}^{Q} &= \frac{1}{2\,\Omega_{pq}}\left( A_p^r A_q^{l*} + A_p^l A_q^{r*}\right) (\nu, \n, t)\\
	B_{pq}^{U} &= \frac{-i}{2\,\Omega_{pq}}\left( A_p^r A_q^{l*} - A_p^l A_q^{r*}  \right) (\nu, \n, t)\\
	B_{pq}^{V} &= \frac{1}{2\,\Omega_{pq}}\left( A_p^r A_q^{r*} - A_p^l A_q^{l*} \right) (\nu, \n, t)
\end{split}
\end{align}
are ``beam transfer functions". For a well-designed optical system with low cross-polarization response (i.e. $A_r^r(\n) \gg A_r^l(\n)$ and $A_l^l(\n) \gg A_l^r(\n)$ over most of the sky), we have
\begin{align}
\begin{split}
	B_{rr}^T(\n), B_{rr}^V(\n) &\gg B_{rr}^Q(\n), B_{rr}^U(\n)\\
	B_{rl}^Q(\n), B_{rl}^U(\n) &\gg B_{rl}^T(\n), B_{rl}^V(\n)\\
	B_{ll}^T(\n), B_{ll}^V(\n) &\gg B_{ll}^Q(\n), B_{ll}^U(\n).
\end{split}
\end{align}
It can be shown that any optical design with continuous azimuthal symmetry that is excited with circular polarization in an $m=1$ waveguide mode (like the fundamental $TE_{11}$ mode) has radiation patterns that take the form \cite{kildal_textbook}
\begin{alignat}{4}\label{eq:circ_sims_radiation_patterns_repeated}
	&A_l^{l}(\nu, \n) = &&A_{\text{co}}(\nu, \theta)e^{i\phi}, \quad&&A_l^{r}(\nu, \n) = &&A_{\text{cx}}(\nu, \theta)e^{i\phi}\nonumber\\
	&A_r^{l}(\nu, \n) = -&&A_{\text{cx}}(\nu, \theta)e^{-i\phi}, \quad&&A_r^{r}(\nu, \n) = -&&A_{\text{co}}(\nu, \theta)e^{-i\phi}.
\end{alignat}
$A_{\text{co}}(\nu, \theta)$ and $A_{\text{cx}}(\nu, \theta)$ are complex-valued functions that, at a given frequency, only depend on the angle from the boresight and their form depends on the detailed geometry of the azimuthally symmetric system. $A_{\text{co}}(\nu, \theta)$ describes the co-polarization response of the antenna and $A_{\text{cx}}(\nu, \theta)$ describes the cross-polarization response.

Since we measure $Q$ and $U$ from the cross-correlation product, it is worth writing out its form for the hat feed explicitly
\begin{align}
	W_{rl}(t) = \frac{1}{2\Omega_{rl}}\int d^2\n \, &\big[-2\,{\rm Re} \big(A_{\rm co}(\theta)A_{\rm cx}^*(\theta)\big)e^{-2i\phi}\,T(\n,t) - \left(|A_{\rm co}(\theta)|^2 + |A_{\rm cx}(\theta)|^2\right)e^{-2i\phi}\,Q(\n, t) \nonumber\\ & + i\left(|A_{\rm co}(\theta)|^2 - |A_{\rm cx}(\theta)|^2\right)e^{-2i\phi}\,U(\n,t) -2i\,{\rm Im} \big(A_{\rm co}(\theta)A_{\rm cx}^*(\theta)\big)e^{-2i\phi}\,V(\n,t)\big]\label{eq:W_rl_hat_feed},
\end{align}
and note that we have changed the time dependence to be on the sky terms to indicate that we are now working in the telescope coordinate frame. $V \simeq 0$ over most of the sky at CMB and CGEM observing frequencies. A minor exception to this is circularly polarized emission from Zeeman splitting of the magnetic dipole transitions of molecular oxygen in the atmosphere \cite{petroff_class, padilla_class}. This emission is expected to be low from 8-10\,GHz and could be removed as a dipole in Stokes $V$ oriented to the Earth's magnetic poles \cite{petroff_class}. Accordingly, we can ignore the Stokes $V$ term for our purposes. Recall also that $|A_{\rm co}(\theta)|^2 \gg |A_{\rm cx}(\theta)|^2$ for well-designed optics. We can think of the $Q$ and $U$ terms as spin-2 convolutions of $Q$ and $U$ with the circularly symmetric beams $(|A_{\rm co}(\theta)|^2 + |A_{\rm cx}(\theta)|^2)/2$ and $(|A_{\rm co}(\theta)|^2 - |A_{\rm cx}(\theta)|^2)/2$, respectively. This means that without the presence of the $T$ and $V$ terms and considering only beam effects, we would recover exact $Q$ and $U$ maps at the angular resolution of our instrument, with only minor systematic leakage between $Q$ and $U$ (which arises from the difference in sign of the $|A_{\rm cx}|^2$ terms in the above expression). This interpretation of the $Q$ and $U$ terms allows us to see that $Q_{\text{measured}} = -\rm 2Re (W_{rl})$ and $U_{\text{measured}} = 2\rm Im (W_{rl})$. Recalling that $T \gg Q,U$ over most of the sky from 8-10\,GHz, we can see that the $T$ term
\begin{align}
	\Delta W_{rl}(t) = \Delta W_{lr}^*(t) \equiv \frac{1}{\Omega_{rl}}\int d^2\n \,  \big[-2\,{\rm Re} \big(A_{\rm co}(\theta)A_{\rm cx}^*(\theta)\big)e^{-2i\phi}\,T(\n,t) \big]\label{eq:rl_leakage_term}
\end{align}
is our most important beam/optics-related systematic, \TtoQU\ leakage. We can also see that $T$ leaks into $Q_{\text{measured}}$ and $U_{\text{measured}}$ in approximately equal amounts (and hence leaks into $E_\text{measured}$ and $B_\text{measured}$ in equal amounts) due to the $e^{-2i\phi}$ term. We therefore concern ourselves with reducing \TtoP\ leakage during the optical design, where $P = \sqrt{Q^2 + U^2}$ is linearly polarized intensity. We also choose to minimize \TtoP\ leakage instead of $T\rightarrow B$ leakage because $\Delta P$ a robust measure of polarization leakage, it is a scalar quantity instead of a spin-2 quantity (and is hence better behaved when taking power spectra on a masked sky), and minimizing $\Delta P$ leakage is guaranteed to minimize $T\rightarrow B$ leakage.

To maximize polarization purity, we would like to minimize \TtoP\ leakage in simulated time-ordered data or sky maps during an optimization. However, this is too computationally expensive to do in an optimization. We can make it tractable to directly minimize these simulated systematics if we do the computation in spherical harmonic space and with a pre-computation step.

\subsubsection{$T\rightarrow P$ Leakage in Spherical Harmonic Space}
Recalling that $T(\n)$ and $B_{pq}^T(\n)$ are spin-0 and taking a spherical harmonic expansion of Equation \ref{eq:rl_leakage_term}, we get
\begin{align}
	\Delta W_{rl}(t) =  \int d^2\n \, \left[\left(\sum_{\ell m} b_{\ell m} Y_{\ell m}(\n)\right)\left(\sum_{\ell' m'} t_{\ell' m'}(t) Y_{\ell' m'}(\n)\right)\right]\label{eq:initial_sph_harm_trans}
\end{align}
where
\begin{align}
	&b_{\ell m} = \frac{1}{{\Omega_{rl}}} \int d^2\n\, \left(-2\,{\rm Re} \left(A_{\rm co}A_{\rm cx}^*\right)e^{-2i\varphi} \right)Y^*_{\ell m}(\n)\label{eq:blm_starting_point}, \quad t_{\ell m}(t) = \int d^2\n\, T(\n, t) Y^*_{\ell m}(\n).
\end{align}
Note that because $T(\n)$ is real, $T(\n) = T^*(\n)$.
We can use this to simplify Equation~\ref{eq:initial_sph_harm_trans}:
\begin{align}
	\Delta W_{rl}(t)  &= \sum_{\ell m}\sum_{\ell'm'} b_{\ell m} t^*_{\ell'm'}(t)\int d^2\n \, Y_{\ell m}(\n)Y^*_{\ell'm'}(\n) = \sum_{\ell m} b_{\ell m} t^*_{\ell m}(t),
\end{align}
where we have used the orthogonality of the spherical harmonics to collapse the integral and eliminate one of the sums. From Equation \ref{eq:blm_starting_point}, it can be shown that
\begin{align}
		b_{\ell m} &= -\frac{4\pi}{\Omega_{rl}}\sqrt{\frac{2l + 1}{4\pi} \frac{(l - 2)!}{(l + 2)!}}\int^\pi_0 d\theta \sin\theta\,{\rm Re} \left(A_{\rm co}(\theta)A_{\rm cx}^*(\theta)\right) P_\ell^{2}(\cos \theta) \delta_{m, -2}\label{eq:b_lm}.
\end{align}
The leakage (at a single time sample) can therefore be written as
\begin{alignat}{2}
	\Delta W_{rl}(t) &= \sum_{l} \Bigg[-\frac{4\pi}{\Omega_{rl}}\sqrt{\frac{2l + 1}{4\pi} \frac{(l - 2)!}{(l + 2)!}} \int^\pi_0 d\theta \sin\theta\,{\rm Re} \left(A_{\rm co}(\theta)A_{\rm cx}^*(\theta)\right) P_\ell^{2}(\cos \theta)\Bigg] t_{\ell,-2}^*(t) \nonumber\\
	&= \sum_{l} b_{\ell,-2} t^*_{\ell,-2}(t).\label{eq:corr_leakage}
\end{alignat}
We can see that the leakage is encoded in a small number of non-zero $a_{\ell m}$'s and this sum-product can be computed rapidly once $b_{\ell,-2}$ and $t^*_{\ell,-2}$ are known. $b_{\ell, -2}$ can be computed from the radiation patterns of a candidate optical design in a fraction of a second.

Since $Q_{\text{measured}} = -2\rm Re(W_{rl})$ and $U_{\text{measured}} = \rm 2Im(W_{rl})$, the leakage of $T$ into $Q$ is $\Delta Q = -\rm Re(\Delta W_{rl})$ and the leakage of $T$ into $U$ is $\Delta U = \rm Im(\Delta W_{rl})$ (and recall that we already absorbed a factor of 2 into the definition of $\Delta W_{rl}$ in Equation ~\ref{eq:rl_leakage_term}). The \TtoP\ leakage is then
\begin{align}
	\Delta P(t) &= \sqrt{\Delta Q(t)^2 + \Delta U(t)^2} = \sqrt{\left(-\text{Re}\left[\Delta W_{rl}(t)\right]\right)^2 + \left(\text{Im}\left[\Delta W_{rl}(t)\right]\right)^2} = \left| \Delta W_{rl}(t) \right| = \left| \Delta W_{lr}(t) \right|.\label{eq:delta_P_simplified}
\end{align}
Therefore, a suitable optimization goal is to minimize  $\Delta P(t) = | \Delta W_{rl} | = |\sum_{l} b_{\ell,-2} t^*_{\ell,-2}(t)|$, which will minimize $T\rightarrow P$ leakage and will maximize polarization purity.

\subsubsection{Fast Simulations of $T\rightarrow P$ Leakage}\label{sec:incorporating_scan_strategy}

Recall that we are working in the telescope frame. In order to evaluate and minimize $\Delta P(t)$, we need $t^*_{\ell,-2}(t)$, which changes as a function of time as the telescope scans and the sky moves overhead. Given the scan strategy for CGEM and a model for $T(\n)$ near $9\,$GHz, we compute the $t^*_{\ell,-2}$ in the celestial frame and rotate the $t^*_{\ell,-2}$ into the telescope frame using DUCC \cite{ducc}. We do these rotations for a complete set of pointings for a given scan strategy: one rotated set of $t^*_{\ell,-2}$ for each observation of each pixel with a unique telescope orientation. For example, each pixel is observed with $\sim$2 unique telescope orientations for the CGEM constant-elevation scan strategy, and so we get one set of $t^*_{\ell,-2}$ for each of $2\times N_{\text{pix}}$ unique pointings. The $t^*_{\ell,-2}$ rotation is a cheap, embarrassingly parallel computation that only needs to be done once for a scan strategy and model sky.  For CGEM, we used a scaled version of the Remazeilles-reprocessed Haslam $408\,$MHz synchrotron map \cite{haslam, res_haslam} as our model for $T(\n)$, scaled to 8-10\,GHz using a spatially-independent synchrotron scaling relation $T_\nu \simeq T_{408\,\text{MHz}}(\nu\,\text{MHz}/408\,\text{MHz})^{-2.7}$ \cite{unpol_scaling_faraday}.

Using the precomputed set of $t^*_{\ell,-2}(t)$, we can quickly compute $\Delta P(t) = |\sum_{l} b_{\ell,-2} t_{\ell,-2}(t)|$ as follows:
\begin{itemize}
	\item Given an optical design geometry, evaluate the radiation patterns in TICRA Tools to obtain $A_{\rm co}(\theta)$, $A_{\rm cx}(\theta)$.
	\item Compute $b_{l,-2}$ from $A_{\rm co}(\theta)$, $A_{\rm cx}(\theta)$ using Equation \ref{eq:b_lm}.
	\item Using the set of precomputed $t^*_{\ell,-2}(t)$ and $b_{\ell,-2}$, compute $\Delta P(t) = |\sum_{l} b_{\ell,2} t_{\ell,2}(t)|$ to get the $T\rightarrow P$ leakage for each unique telescope pointing.
\end{itemize}
On a single computing node with 16 cores, computing $\Delta P(t)$ from the last two steps takes about 10 seconds at NSIDE 256 ($\ell_{\text{max}} = 767$), and the runtime of the three steps together is dominated by the TICRA simulation step (which takes about 1 minute for the CGEM optics). We can then make a loss function from $\Delta P(t)$ to use in an optimization to adjust the optical design to minimize our most important beam-related polarization systematic. The key point is that a $\Delta P$ loss function value can be computed fast enough for practical use in numerical optimization algorithms (e.g. gradient descent). The pre-computation of $t_{\ell,-2}(t)$ took 22\,h, but this only had to be done 6 times for the entire CGEM optical design (for constant-elevation scan strategies at three different elevations for two different sky models).

\subsubsection{Hat Feed Parameterization and Initial Geometry}\label{sec:opt_goals_and_hat_feed_params}

We parameterized the hat feed as follows to allow its geometry to be adjusted in an optimization:
\begin{itemize}
    \item \textbf{Feedhorn:} Following Ref.~\citenum{smooth_walled_horn}, we use a smooth-walled feedhorn with a parameterized, monotonically increasing spline profile to achieve excellent optical performance and to have a horn that is inexpensive to manufacture. The horn profile is parameterized by 40 nodes which are evenly spaced between the throat and aperture of the horn, are monotonically increasing in radius along the length of the horn, and are interpolated with a Piecewise Cubic Hermite Interpolating Polynomial (PCHIP), which preserves monotonicity between the nodes. The radii of each node from the centre of the horn are parameters. We therefore have 41 parameters to represent the horn: the length of the horn, and the radii of the 40 nodes composing the horn profile.
    \item \textbf{Hat Feed Geometry:} The ring-focus hat feed geometry described in Section~\ref{sec:hat_feed_intro} can be uniquely determined by 4 ``principal" parameters: the diameters of the primary and secondary mirror, the focal length of the primary mirror, and the distance of the secondary mirror from the feed aperture. We keep the primary mirror diameter fixed to 4\,m and the focal length fixed to 1\,m, but allow the secondary mirror diameter and distance from the horn aperture to vary. This preserves a ``deep-dish" design with less spillover past the primary mirror.
    \item \textbf{Reflector Surfaces:} As previously mentioned, we allow the optimizer to add perturbations to the reflector surfaces. The primary and secondary reflector surfaces are represented by 10 and 20 nodes, respectively, which are interpolated with cubic splines. The positions of the nodes parallel to the optic axis are allowed to vary as parameters.
\end{itemize}

We set the initial horn profile to be same as the initial guess used in Ref.~\citenum{smooth_walled_horn} scaled to 9\,GHz, but shortened in length by 80\%. We initialized the hat feed geometry such that the secondary mirror was $\sim$30\,cm in radius and $\sim$10\,cm from the horn aperture. We set the initial surface perturbations to zero.

This parameterization results in 75 total parameters. Note that in reality, the optics were optimized in several stages with different optimization goals and parameterizations. For example, the primary mirror was manufactured prior to the development of the optimization framework described in this paper, and hence its shape was determined with different optimization goals and parameters. However, in principle, all of these parameters could be optimized at once with this framework and we rarely found the optimizer to struggle with high numbers of parameters.

\subsubsection{Optimization Goals and Procedure}\label{sec:opt_goals}

We use three optimization goals: a) maximize the forward gain (to improve our angular resolution), b) minimize the $S_{11}$ of the horn and reflectors measured at the throat of the horn, and c) minimize the power spectrum of $\Delta P$ on the large angular scales relevant to the \bmode\ search. Explicitly, the loss function that we minimize is
\begin{align}
	f = w_G\sum_\nu G_\nu^2 + w_S\sum_\nu S_\nu^2 + w_P\sum_{\nu} \sum_{\ell = 2}^{120} \frac{\ell(\ell + 1)}{2\pi}C_{\nu,\ell}^{\Delta P \Delta P},\label{eq:loss}
\end{align}
where $G_\nu = \text{max}\left\{\text{Forward Gain}(\nu) - 50\,\text{dBi}, 0 \right\}$, $S_\nu = \text{max}\left\{S_{11}(\nu) - (-20\,\text{dB}), 0\right\}$ and $w_G$, $w_S$, and $w_F$ are weight factors that are fixed at the start of the optimization to control the relative importance of each goal. The leakage goal with $C_{\nu, \ell}^{\Delta P \Delta P}$ is described in more detail below. We empirically found that we get the best optimized performance when $w_S$ and $w_P$ are such that the corresponding goals contribute 16$\times$ as much to the loss as the forward gain goal. If the leakage goal was not upweighted, it was found that the optimizer would fine-tune the forward gain and $S_{11}$ goals at the cost of higher leakage levels. If the $S_{11}$ goal was not upweighted, the optimizer would have trouble finding solutions that had $S_{11} < -20\,$dB, particularly when the initial guess for the horn length was shortened. The final weights merely reflect how much we care about each goal in the context of CGEM.

One evaluation of the loss function proceeds as follows:
\begin{itemize}
	\item Given a set of optical design parameters, call TICRA Tools from a Python program to evaluate the radiation patterns and $S_{11}$ for the corresponding optical design at 50 evenly-spaced frequencies between 8 and 10\,GHz.
	\item Extract the forward gain in dBi from the TICRA Tools radiation patterns and evaluate $G_\nu$. Then extract $S_{11}$ from the TICRA Tools output and evaluate $S_\nu$.
	\item Evaluate $\Delta P_\nu$ for each unique pointing in the scan strategy using the procedure described at the end of Section \ref{sec:incorporating_scan_strategy}.
	\item Bin the $\Delta P_\nu$ values into HEALPix maps ($\Delta P_\nu(\n)$), mask the galaxy for $|b| < 20^\circ$ and unobserved regions in the maps, subtract the monopole and dipole of each map, apodize the maps near the mask edges, and compute their power spectra, defined as
    \begin{equation}
	C_{\nu, \ell}^{\Delta P\Delta P} = \frac{1}{2\ell + 1}\sum_m a_{\nu, \ell m}^{\Delta P*} a_{\nu, \ell m}^{\Delta P}, \quad a_{\nu, \ell m}^{\Delta P}  = \int d\n\, Y_{\ell m}^*(\n) \Delta P_\nu(\n),
\end{equation}
with HEALPix's anafast routine. 
	\item Sum $\ell(\ell+1)C_{\nu, \ell}^{\Delta P\Delta P}/(2\pi)$ over the $\ell$ range: $2 < \ell < 120$. We use this $\ell$ range to focus polarization purity improvements on the large scales relevant to the B-mode search. 
	\item Sum each of the loss terms over frequency and add them together to get the total loss.
\end{itemize}
We used SciPy's trust-constr routine \cite{scipy} to minimize the loss function. We added constraints to the optimization to ensure that the horn nodes are monotonically increasing along its length. We also added upper and/or lower limits on each parameter to, for example, prevent the hat diameter and horn aperture from getting too large. We monitored the total loss and the loss for each goal throughout the optimization and terminated the program when the loss had not decreased significantly for several thousand iterations.

It is important to note that we re-ran the optimization with different scan strategies and model skies, and found the resulting optimized geometry and polarization purity to be insensitive to these choices, as long as a somewhat realistic Galactic plane was included in the model sky. It is also important to note that we have ignored the effect of the ground screen in this $T\rightarrow P$ leakage computation. However, we have simulated $T\rightarrow P$ leakage angular power spectra for optics that include and do not include a ground screen using a custom convolution code. We repeated these simulations with and without ground emission. When ground emission is not included, we find that the ground shield, which predominantly alters far sidelobe structure in the beam where the response is low, does not significantly change the $T\rightarrow P$ leakage angular power spectra. When ground emission is included, the $T\rightarrow P$ leakage is much worse for the beam that does not include the ground shield. We therefore argue that it is justified to ignore the effect of the ground shield in our optimization framework, and that the net effect of the ground shield is to greatly reduce $T\rightarrow P$ leakage.

\subsection{Optimization Results}

\subsubsection{Optimized Geometry}
\begin{figure} [ht]
   \begin{center}
   \begin{tabular}{c} 
   \includegraphics[height=7cm]{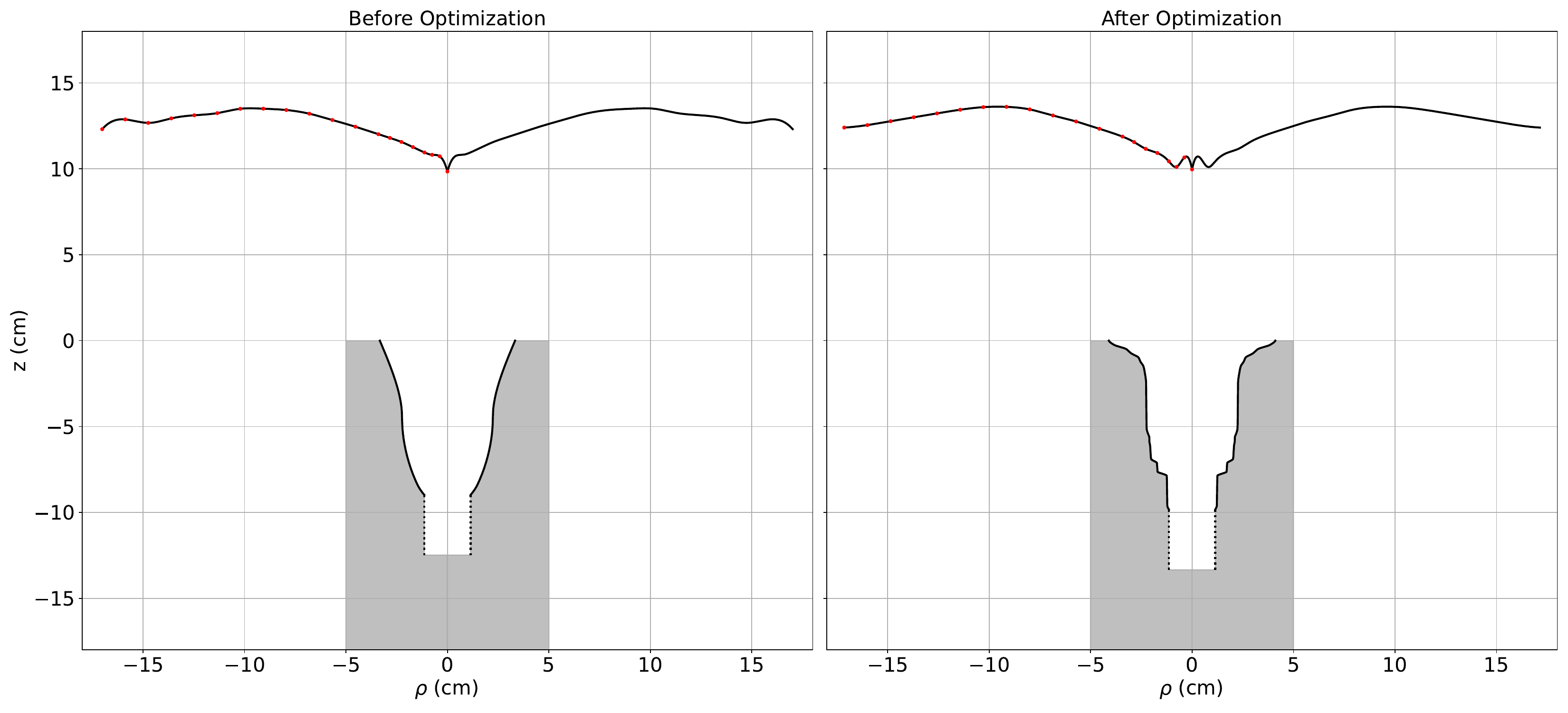}\\
   \includegraphics[height=7cm]{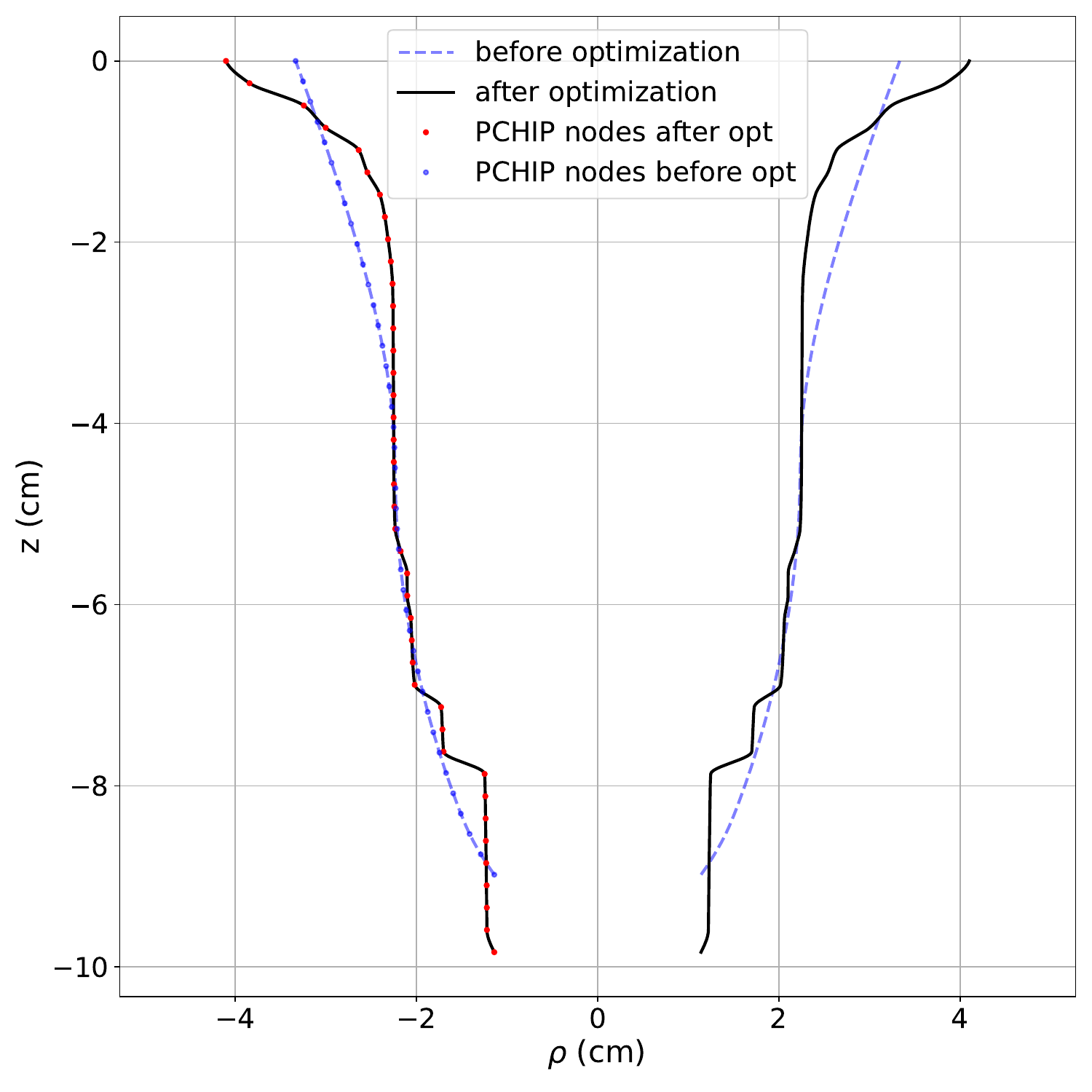}
   \end{tabular}
   \end{center}
   \caption[example] 
   {Top left: the horn and hat geometry prior to the final optimization that produced the CGEM optical geometry. As mentioned in Section~\ref{sec:opt_goals_and_hat_feed_params}, the optimization proceeded in several stages prior to the development of the optimization framework described in Section~\ref{sec:opt_framework}. The hat geometry shown in this panel was the result of an intermediate optimization. Top right: the horn and hat geometry after the final optimization. This is the geometry of the currently-deployed CGEM optics. Bottom: a closeup of the horn profile before and after optimization. In all panels, red dots represent the shape-defining spline nodes that were adjusted in the optimization. The dashed lines in the top two panels at the throat of the horn illustrate the input waveguide. We can see that the optimizer made significant changes to the horn profile and less drastic changes to the hat. \label{fig:before_after_opt_geometry}}
\end{figure}

Figure~\ref{fig:before_after_opt_geometry} shows the geometry of the horn and hat before and after the final optimization that produced the CGEM optical geometry. As mentioned in Section~\ref{sec:opt_goals_and_hat_feed_params}, the optimization proceeded in several stages prior to the development of the optimization framework described in Section~\ref{sec:opt_framework}. The hat geometry shown in the left panel (before optimization) was the result of an intermediate optimization. The primary mirror is not shown in Figure~\ref{fig:before_after_opt_geometry} because the optimized shape only differs from the initial shape by at most $\sim$2\,cm. The upper left and right panels show the horn and hat geometry on an equal scale before and after optimization, respectively. The bottom panel shows a close-up of the horn profile before and after optimization. We can see that the horn was heavily shaped during the optimization, while the changes to the hat were less drastic. There was a large reduction in the loss for each goal when the optimizer added the straight segment of the horn just after the throat. The central dips and sharp point on the hat were empirically observed to improve the $S_{11}$ of the optics. Manually reducing the size of the sharp point or dulling it in simulations was found to not significantly impact performance. Despite being heavily shaped, the horn and hat features are simple enough to be made on a computer numerical control (CNC) lathe, and this was partly enforced by the parameterization (e.g. the monotonicity constraint on the horn profile). 

\subsubsection{Standard Antenna Figures of Merit}
\begin{figure} [ht]
   \begin{center}
   \begin{tabular}{c} 
   \includegraphics[width=14cm]{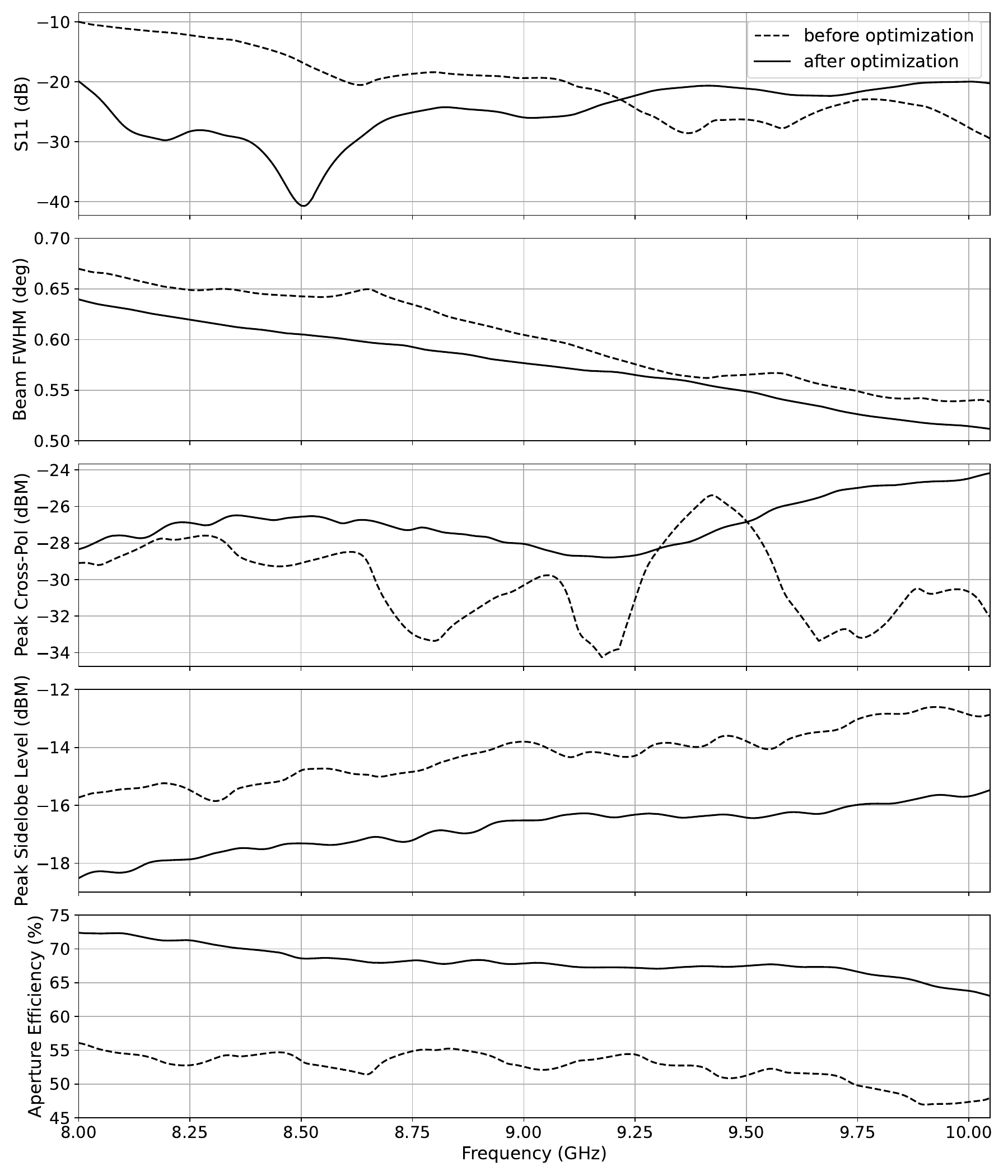}
   \end{tabular}
   \end{center}
   \caption[example] 
   { \label{fig:opt_87_standard_beam_params} 
From top to bottom: 1) $S_{11}$ measured at the throat of the horn (including the effect of both reflectors), 2) the beam full-width at half-maximum, 3) the peak cross-polarization response relative to the peak co-polarization response, 4) the peak co-polarization sidelobe level relative to the peak of the main beam, 5) the aperture efficiency of the optics. These standard antenna figures of merit suggest that the optimized design has good performance and that the performance was improved in the optimization.}
\end{figure}

Figure~\ref{fig:opt_87_standard_beam_params} shows several standard metrics for antenna performance evaluated for the hat feed optics before and after optimization. The top two panels show $S_{11}$ and beam width (inversely proportional to forward gain), which were figures of merit in our optimization. We can see that $S_{11}$ is below $-20$\,dB (1\%) over the whole band and that the beam width is near to 0.5$^\circ$, as desired. The peak cross-polar response relative to the co-polar peak is plotted in the middle panel. Interestingly, the peak cross-polar response has increased slightly during the optimization, despite the optimized design having lower \TtoP\ leakage (as we will see in the next section). Also shown in Figure \ref{fig:opt_87_standard_beam_params} are the peak co-polarization sidelobe level and aperture efficiency. The co-polarization sidelobe level and aperture efficiency have clearly improved during the optimization, even though they were not being explicitly optimized. Note that we achieve a fairly high aperture efficiency (65-70\%), even though the hat feed system has a partially blocked aperture. These standard antenna performance metrics suggest that the optimized design has good performance.

\subsubsection{$T\rightarrow P$ Leakage}
\begin{figure} [ht]
   \begin{center}
   \begin{tabular}{c} 
   \includegraphics[height=12cm]{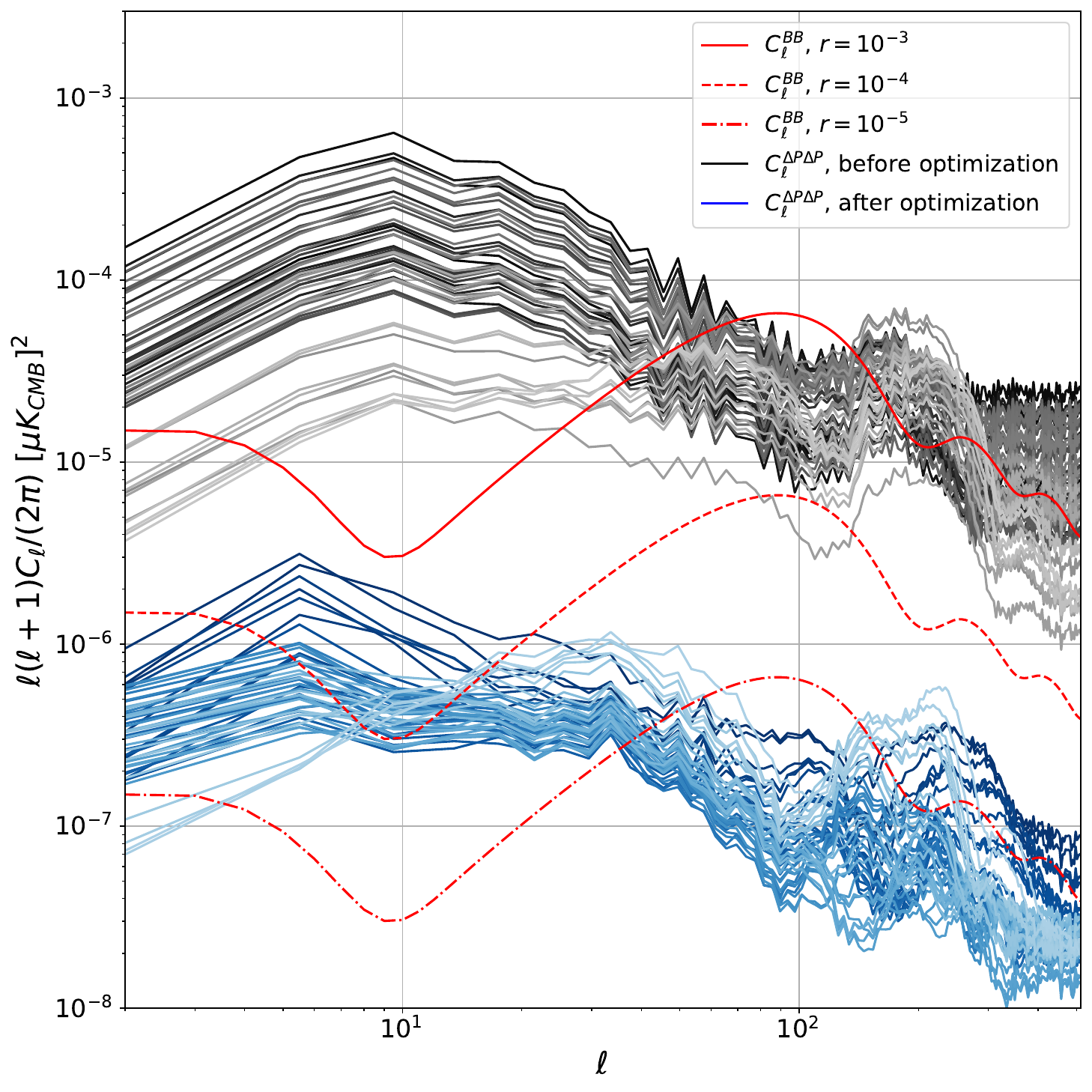}
   \end{tabular}
   \end{center}
   \caption[example] 
   {Simulated $\Delta P$ angular power spectra for the hat feed optical design before and after optimization. The groups of $\Delta P$ power spectra are shown at 50 frequencies from 8-10\,GHz, where each curve has been scaled to the CMB observing window at 95\,GHz (assuming a synchrotron spectral index of $-3.1$) and converted to CMB thermodynamic units. Power spectra are averaged in $\ell$ bins of size 4. The darkest shade in each colour family corresponds to 8\,GHz (scaled to 95\,GHz), while the lightest shade corresponds to 10\,GHz (scaled to 95\,GHz). \bmode\ power spectra for three different values of $r$ are also plotted (with no contribution from lensing B-modes). The optimized optical design has $\sim$1-2 orders of magnitude less leakage at all angular scales compared to the initial design, and the leakage is below $r=10^{-3}$ by $\sim$1-2 orders of magnitude.\label{fig:leakage_p_spec_before_after_opt}}
\end{figure}

The most important diagnostic of polarization purity for our science goals is the level of the systematic \TtoP\ leakage angular power spectra compared to theoretical \bmode\ power spectra. Figure \ref{fig:leakage_p_spec_before_after_opt} shows simulated $\Delta P$ angular power spectra for the hat feed optical design before and after optimization with the framework in Section \ref{sec:opt_framework}. These power spectra are minimized in the optimization for $\ell < 120$. To reiterate from Section \ref{sec:opt_goals}, these are produced by taking $\Delta P$ maps, masking the galaxy for $|b| < 20^\circ$ and unobserved regions, apodizing the masked maps near the mask edges, and taking the power spectrum with the HEALPix anafast routine \cite{healpix}. All $\Delta P$ power spectra are scaled from their native CGEM frequencies to the CMB observing window at 95\,GHz assuming a polarized synchrotron spectral index of $-3.1$, and have been converted to CMB thermodynamic units. The darkest shade in each colour family corresponds to 8\,GHz (scaled to 95\,GHz), while the lightest shade corresponds to 10\,GHz (scaled to 95\,GHz). The $\Delta P$ power spectra are averaged into $\ell$ bins of size 4. We also plot primordial \bmode\ power spectra (with no lensing B-modes) for three different values of the tensor to scalar ratio, $r$. The next generation of CMB experiments are targeting upper limits of $r \lesssim 10^{-3}$ \cite{cmb_s4_science_book}. We can see that before the optimization, the $\Delta P$ angular power spectra are comparable to $r=10^{-3}$. After the optimization, the leakage is reduced by 1-2 orders of magnitude and is far below $r=10^{-3}$. This suggests that the optics have high polarization purity and excellent performance for CGEM to aid in the \bmode\ search. This also demonstrates the power of this optimization framework to directly minimize polarization systematics in the space that is most relevant to our science goals.

\section{Mechanical Design}\label{sec:mech}

\begin{figure} [ht]
   \begin{center}
   \begin{tabular}{c} 
   \includegraphics[height=7.25cm]{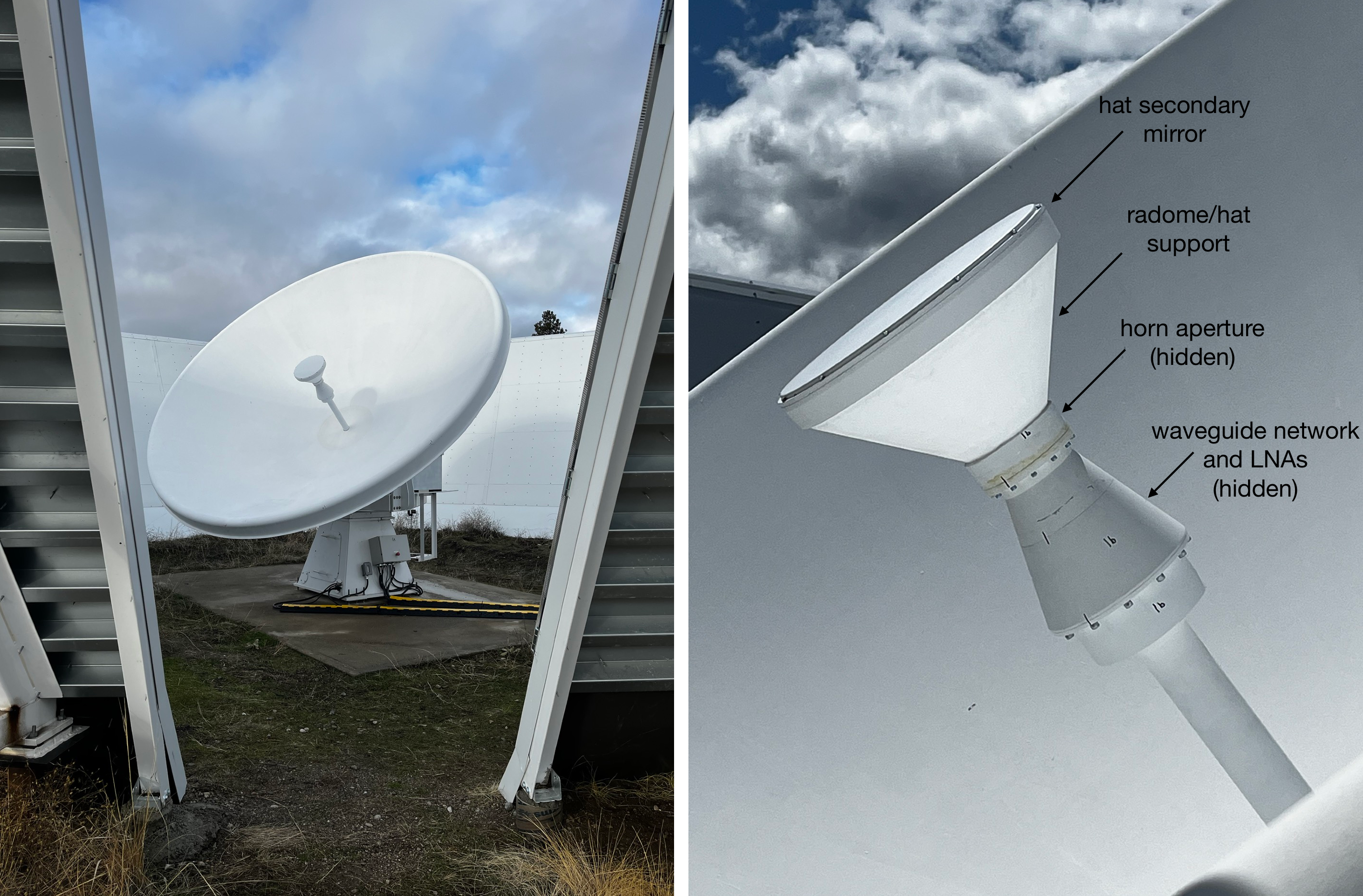}
   \includegraphics[height=7.25cm]{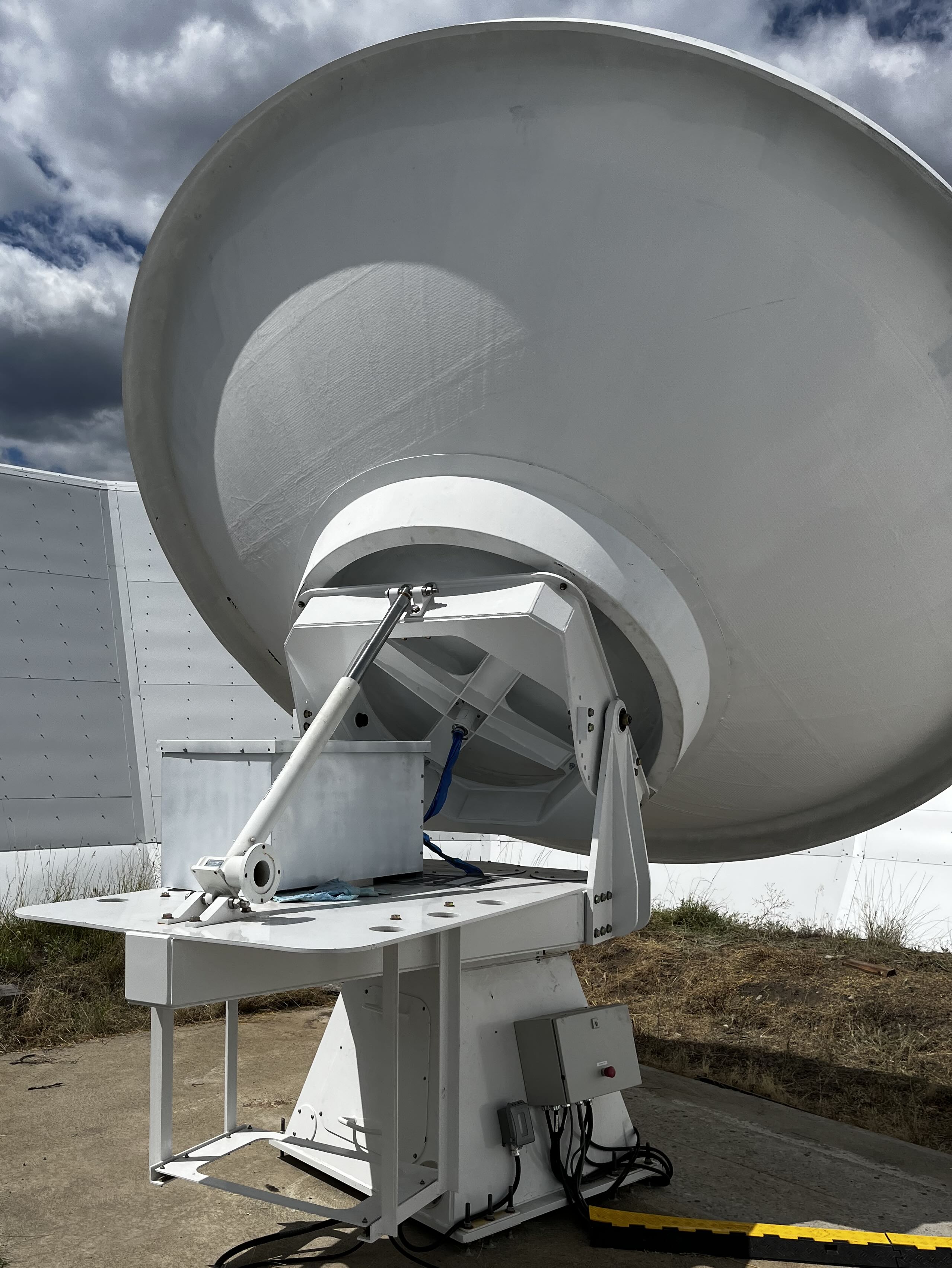}
   \end{tabular}
   \end{center}
   \caption[example] 
   {Images of the CGEM optics as they are currently deployed. Left panel: a face-on view of the optics during an azimuth scan. Middle panel: a close-up of the feed column and hat support, with key components labeled. Right panel: a view of the back of the dish to illustrate the ring backing structure on the primary mirror, and the steel cross with set screws that holds the feed column in place. The tree tip visible over the ground shield in the photo at left is not visible from the location of the telescope.
   \label{fig:deployed_optics}}
\end{figure}

Here we describe the mechanical design of the optics. Great care was taken to design each component to realize the optimized performance shown in the previous section. In the following subsections, we describe the design and manufacturing of each component in turn.

\subsection{Primary Mirror}\label{sec:primary_mirror_mech}
The primary mirror was manufactured by the Composite Applications for Radio Telescopes (CART) team at the DRAO. This team specializes in making lightweight reflectors out of composite materials. The manufacturing process involved laying up several layers of carbon fibre and a reflective copper mesh layer on a mould with the desired shape, vacuum bagging the mould and materials, injecting resin, evacuating the vacuum bag to evenly distribute the resin, and putting the dish and mould in an oven to cure the resin and solidify the dish shape. The result is a lightweight ($\sim$100\,kg) and strong dish with a very accurate surface shape. A ``ring" backing structure is also bonded to the back of the dish and this has mounting holes to fasten the dish to the CGEM mount. The CART team made 3 dishes: one for CGEM, one for a future Southern-hemisphere copy of CGEM, and a spare dish. The dish surface shapes were measured by bolting them to a platform and measuring deviations from the design shape with a laser tracker and track ball. The surface errors are less than $\sim$125$\,\mu$m RMS. The measured shapes were imported into TICRA Tools and surface errors were found to negligibly impact the simulated performance.

A finished dish deployed on the telescope mount can be seen in the left panel of Figure \ref{fig:deployed_optics}, while the backing/mounting structure can be seen in the right panel.

It was found that if the bolts to fasten the dish to the telescope mount were tightened in different orders and with different torques, this could induce a small but measurable saddle-shaped ($m=2$) distortion in the dish surface. For this reason, the CGEM primary mirror must be measured and adjusted after being deployed on the telescope mount (we will see the need for this in Section \ref{sec:misalignments_and_model}). 

\subsection{Feedhorn}\label{sec:horn}
The smooth-walled feedhorn was manufactured by Spin Industries Ltd in Vancouver out of aluminium 6061 on a CNC lathe. 3.5\,cm of additional waveguide was added at the throat to ensure that higher-order modes (which can propagate in the horn but not in the circular waveguide) are killed off before reaching the OMT. The walls are 5\,mm thick throughout. A flange was added at the end of the horn with two rings of bolt holes: an inner ring with holes/slots for pins to allow precise mating with the OMT, and an outer ring which mates to the feed column (see Section \ref{sec:feed_column}). The outer surface of the flange is carefully machined to help with alignment of the horn and hat. A procedure was developed to precisely characterize and measure the $S$ parameters of all devices in our waveguide network, including the feedhorn. The measured $S_{11}$ of the horn agrees very closely with the design performance.

\subsection{Secondary Mirror}\label{sec:hat}
\begin{figure} [ht]
   \begin{center}
   \begin{tabular}{c} 
   \includegraphics[height=8cm]{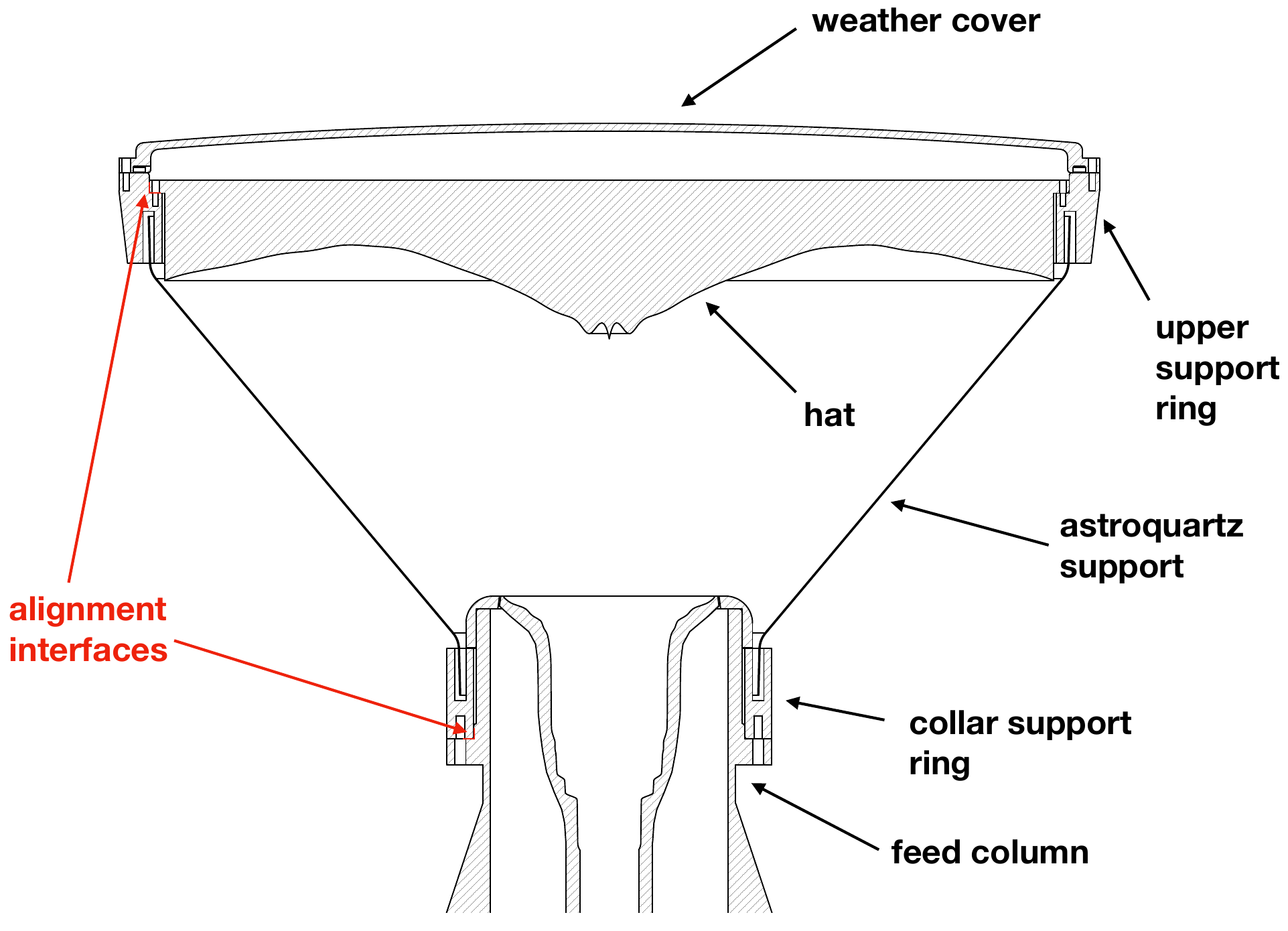}
   \end{tabular}
   \end{center}
   \caption[example] 
   {A cross section of the hat support design, which is a body of revolution. The hat is held by a radome, which consists of two aluminium rings (the upper support ring and collar support ring) glued onto the ends of a cone made out of composite Astroquartz (which has low loss and a low dielectric constant). The radome bolts to a feed column (described in Section~\ref{sec:feed_column}) and holds the hat in front of the horn. A gluing jig for the radome and precision alignment interfaces (shown in red above) ensure that the horn and hat are concentric and have the correct $z$ separation. A weather cover with an O-ring seal ensures that water doesn't enter the volume containing the horn and hat.\label{fig:hat_support_cross_section}}
\end{figure}

The secondary mirror was also manufactured by Spin Industries Ltd out of aluminium 6061 on a CNC lathe. The part has a rim with slightly larger diameter than the reflecting surface. The rim has clearance holes to fasten the hat to the hat support structure (see Section \ref{sec:hat_support}) and a carefully machined outer surface to help with alignment to the horn. The back of the hat is ``pocketed" to reduce its mass while maintaining rigidity. The hat is roughly 5\,mm thick at all points. This part is heavier than is necessary (3.8\,kg) and we will reduce its mass in the design of the optics for the cryogenic version of CGEM. Laser scanner measurements of the hat surface indicate negligible surface errors of $\sim$75\,$\mu \text{m}$ or less. A cross-section of the secondary mirror part can be seen in Figure~\ref{fig:hat_support_cross_section}.

\subsection{Secondary Mirror Support}\label{sec:hat_support}

\subsubsection{Overview}
A cross section of the hat support design, which is a body of revolution, is shown in Figure \ref{fig:hat_support_cross_section}. The hat is held in front of the horn by a support structure that consists of two aluminium rings that are glued to a cone made out of composite material. We refer to the composite and two rings as the ``radome", which is described in Section~\ref{sec:radome}. The horn is held by the feed column (described in Section \ref{sec:feed_column}). The aluminium ring at the narrow end of the radome, called the ``collar support ring", is bolted to a flange on the outside of the feed column. The aluminium ring at the wide end of the radome, called the ``upper support ring", carries the hat secondary mirror (fastened to the ring with bolts) and holds it at the design position in front of the horn. There are two precision-machined alignment interfaces on the rings of the radome to ensure that the hat is placed at the design position in front of the horn and so that the hat and the horn are concentric. The alignment interfaces are shown in red in Figure \ref{fig:hat_support_cross_section}. Also bolted to the upper support ring is an aluminium plate called the ``weather cover". There is an O-ring seal at the interface between the weather cover and the upper support ring that ensures water does not enter the volume containing the horn and the hat. The composite material also provides a seal against water. The upper support ring, collar support ring, and weather cover are all made out of aluminium 6061, which has a good thermal match to the composite radome material. The upper and collar support rings were made by Spin Instustries Ltd. on a CNC lathe, while the weather cover was made by the UBC machine shop on a CNC mill.

\subsubsection{Composite Radome}\label{sec:radome}
We made the radome from Astroquartz III 4503 fabric (a woven silica fabric) pre-impregnated with TC350-1 resin from Toray Advanced Composites. The cured prepreg has exceptional strength and excellent electrical properties: it has a dielectric constant of 3.43 and a loss tangent of 0.023, both at 10\,GHz. Mechanical simulations indicated that 3 or 4 layers of this material ($\sim$0.4\,mm thick) in a conical shape could support the mass of our upper hat support components ($\sim$6.2\,kg) with only a 40\,$\mu$m deflection under gravity when tilted horizontally. At 0.4\,mm thick, a radome with these electrical properties has a noticeable but small effect on the simulated optical performance and \TtoP\ leakage.

The CART team manufactured two radomes: one with 3 layers of Astroquartz and one with 4 layers. To manufacture the radome, the prepreg was laid on a mould with the desired shape, a vacuum bag was placed over the mould and fabric, the bag was evacuated to ensure even distribution of the resin, and the fabric and mould were placed in an oven to cure the resin and give the part its desired shape. The 3- and 4- layer radomes are 0.35\,mm and 0.5\,mm thick, respectively.

The radomes were cut to length and were glued to the aluminium collar and upper support rings with a custom jig. The jig ensures that the collar and upper support rings are concentric and at the right separation from each other during gluing. This in turn ensures that the horn and hat are aligned when the system is deployed. With a coordinate measurement machine, we measured the upper and collar support rings to be concentric within 0.2\,mm for the 3-layer radome, and within 0.5\,mm for the 4-layer radome. The rings on both radomes are witin 0.2\,mm of their design separation parallel to the optic axis. These misalignments are simulated to have a neglible effect on the optical performance.

We measured the stiffness of the 3- and 4-layer radomes by rigidly clamping their bases to a table, pulling on the upper support rings with a spring scale, and reading the movement of the upper support rings with a dial indicator. The upper support rings moved by less than 175\,$\mu$m when pulling on the spring scale with a force of 98\,N, which is 1.6x the force the radome would experience when pointed horizontally and supporting the upper hat components under gravity. From this test, we therefore expected the radomes to exhibit negligible deflection in situ.

\subsection{Feed Column}\label{sec:feed_column}

The feed column was designed to place the horn and hat at the correct position relative to the primary mirror, and to house the OMT and LNAs in a water-tight volume. An image of the deployed feed column can be seen in the middle panel of Figure~\ref{fig:deployed_optics}. Precision alignment interfaces, similar to those shown in Figure~\ref{fig:hat_support_cross_section}, ensure that the horn and collar support ring on the base of the radome are concentric, and hence that the horn is concentric with the secondary mirror. Housing the OMT, LNAs, and output coaxial cables requires a volume with a minimum diameter of 154\,mm, which is why there is a large-diameter section near the middle of the feed column. The exterior of the column was shaped to minimize the impact on the simulated polarization purity, which is why it has a conical shape. The feed column parts were made on a lathe and CNC mill by the UBC machine shop.

The base of the feed column extends through a hole in the primary mirror, where it is held by a sleeve with set screws. The sleeve is bolted to a steel cross that is attached to the telescope mount. The cross and base of the feed column can be seen in the right panel of Figure~\ref{fig:deployed_optics}. The set screws can be adjusted to position the feed column at the design position within the primary mirror. We align the feed column (and hence the horn and hat) to the primary mirror using a laser distance measure with a custom mount that is shaped to the rim of the primary mirror. We place the laser distance measure and mount at different locations around the rim of the primary mirror with the dish pointed at zenith, and adjust the set screws until we get the same reading at every rim location (within the precision of the distance measure, which is roughly $\pm 0.3$\,mm).

\section{On-Sky Performance}\label{sec:on_sky_performance}

We have been commissioning the warm radiometer version of CGEM since February 2025 and have carried out a detailed program to characterize the on-sky performance of the optics. Due to the high noise levels of the warm radiometer, we have primarily used the Sun and bright satellites to measure of the beam. From these measurements, we have developed a beam model, based on TICRA Tools simulations of the optics, that is in outstanding agreement with the data. Here we summarize how this beam model was developed and proceed to showcase the agreement between the data and beam model. We find that the optics are performing very close to their design performance in terms of polarization purity.

\subsection{Measured Misalignments and Beam Model}\label{sec:misalignments_and_model}
Upon initial comparison between observations of the Sun and a forward model for the data based off of electromagnetic simulations of the beam (described in Section~\ref{sec:solar}), it was immediately clear that there were small discrepancies between the data and model, attributable to small optical misalignments. These are listed below with a brief description of how each misalignment was measured: 
\begin{itemize}
    \item \textbf{feed column defocus ($\Delta z_\text{column}$):} The feed column is positioned $\sim$2.5\,mm too close to the primary mirror in the $z$ direction (parallel to the optic axis). This was measured by fitting a TICRA model of the beam to the data, where the $z$ displacement of the feed column was the fit parameter.
    \item \textbf{feed column tilt:} When the dish was tilted from zenith down to $45^\circ$ elevation, the feed column was measured to come out of alignment by $\sim$1\,mm at the location of the horn aperture. The displacement was measured by repeating the feed column alignment measurements described at the end of Section~\ref{sec:feed_column} (using the laser distance measure), but with the dish tilted at $45^\circ$.
    \item \textbf{radome sag:} The radome was found to sag under gravity and come out of alignment by $\sim$2.2\,mm (relative to the horn aperture) when the dish was tilted from zenith to $45^\circ$. This was determined from the same method as the feed column tilt. By comparing the data to electromagnetic simulations, it was found that the $\sim$2.2\,mm radome sag is more consistent with the upper support ring on the radome tilting by $\sim$$0.4^\circ$, rather than translating by $\sim$2.2\,mm.
    \item \textbf{primary mirror saddle distortion:} As described in Section~\ref{sec:primary_mirror_mech}, the primary mirror can be distorted in a saddle shape depending on the torquing of the bolts that fasten it to the mount. By comparing the sum of laser distance measurements from the rim to the feed column on opposite sides of the dish as a function of rim location, we measured the magnitude and orientation of a saddle distortion in the primary mirror. The magnitude is such that the diameter is $\sim$1.5\,mm larger than the design diameter along one axis of the saddle, and is $\sim$1.5\,mm smaller along the perpendicular axis.
\end{itemize}
The beam model that matches the data is generated from a TICRA Tools simulation of the optics that includes these misalignments. As we will see in the following sections, the agreement between the data and model including these misalignments is outstanding. This is strong evidence that the misalignments in the model are present in the real optics. We simulated the effect of these misalignments on \TtoP\ leakage (using a similar method to that described in Section~\ref{sec:incorporating_scan_strategy}) to get a sense of how important each misalignment is to CGEM's science goals. This lets us prioritize which misalignments need to be fixed for the cryogenic version of CGEM.

\begin{figure} [ht]
   \begin{center}
   \begin{tabular}{c} 
   \includegraphics[height=12cm]{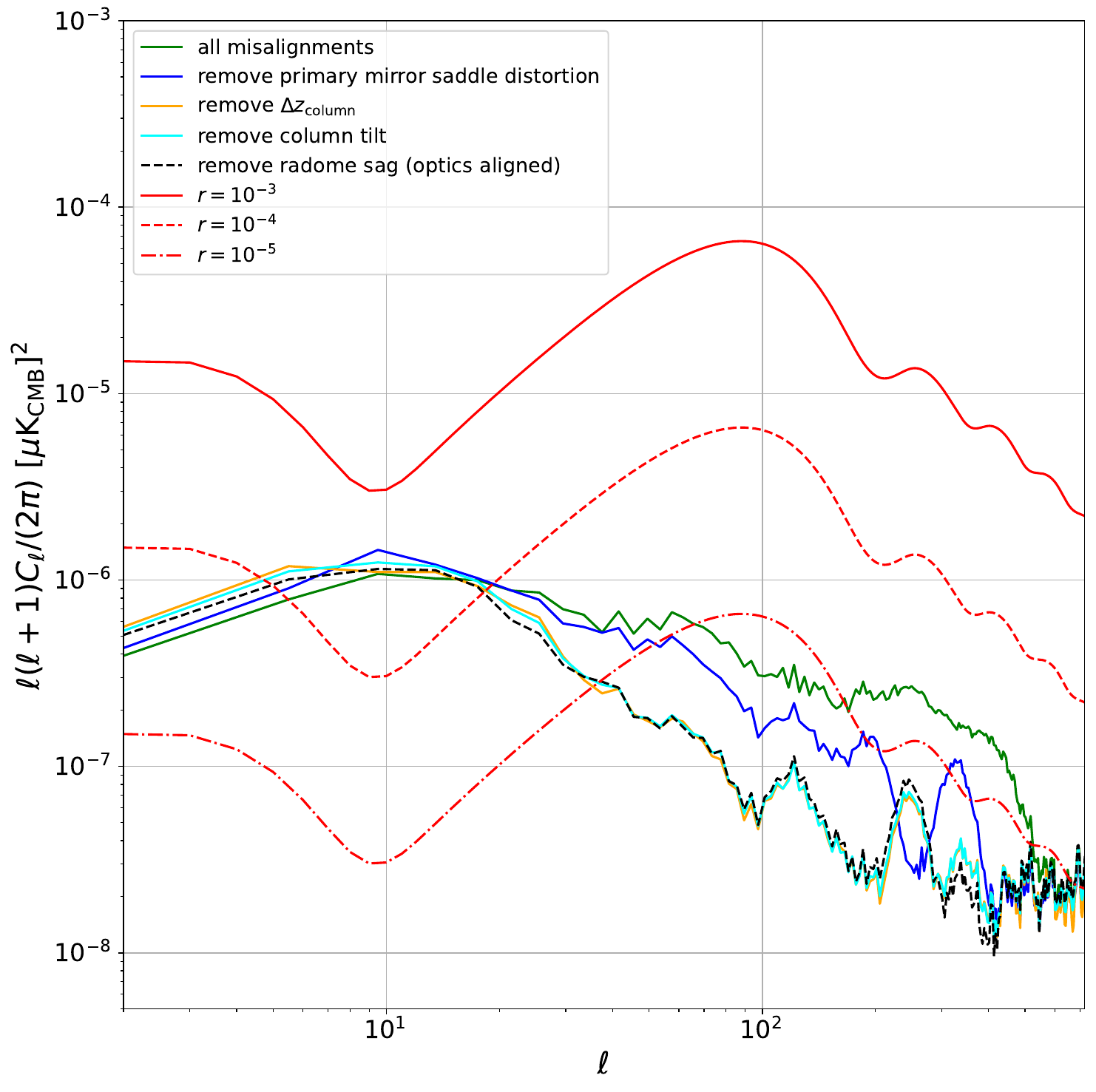}
   \end{tabular}
   \end{center}
   \caption[example] 
   {Simulated \TtoP\ leakage curves for the CGEM optics, where each curve progressively removes a misalignment in the TICRA model. The curves are scaled from 9\,GHz to the CMB observing window at 95\,GHz and are converted to CMB thermodynamic units, as in previous angular power spectrum plots. The plot should be read by looking at the curves in the order that they're presented in the legend (from top to bottom). The green curve shows simulated \TtoP\ leakage for the optical model that includes all known misalignments and best matches the on-sky data. The blue curve contains all the same misalignments but removes the saddle-shaped distortion from the primary mirror, the orange curve then additionally removes the $\Delta z_\text{column}$ misalignment, and so on. We can see that the primary mirror distortion and $\Delta z_\text{column}$ clearly have the biggest impact on leakage and should be corrected first. The leakage, even including all misalignments, is still far below $r=0.001$. This plot is representative of other frequencies in the band.\label{fig:misalign_sims_thesis}}
\end{figure}

Figure~\ref{fig:misalign_sims_thesis} shows simulated  \TtoP\ leakage angular power spectra at 9\,GHz (scaled to 95\,GHz and converted to CMB thermodynamic units) for the different optical misalignments. The black dashed curve shows the \TtoP\ leakage for the perfectly aligned warm radiometer optics. The other curves should be examined in the order that they are listed in the legend (from top to bottom). We start with the optics model in the previous section that has all known misalignments (green curve): $\Delta z_\text{column} = -2.5$\,mm, a feed column tilt such that the horn aperture is displaced by 1\,mm, a 0.4$^\circ$ radome tilt, and a 3\,mm diameter-difference saddle distortion in the primary mirror. The next curve (blue) removes the primary mirror distortion, the next curve (orange) additionally removes the $\Delta z_\text{column}$ misalignment, and so on. We can immediately see that the only misalignments that significantly impact the \TtoP\ leakage are the primary mirror distortion (removing this takes the \TtoP\ leakage from the green curve to the blue curve) and $\Delta z_\text{column}$ (which takes the blue curve to the orange curve). These plots were made at several frequencies throughout the band and we reach the same conclusions about the misalignments. Fortunately, the primary mirror distortion and $\Delta z_\text{column}$ are the easiest to fix: the CART team has a procedure for adjusting the dish to remove saddle-shaped distortions and we can remove $\Delta z_\text{column}$ by adjusting the feed column set screws and shimming the column. The radome sag is the hardest misalignment to fix and fortunately, this hardly has any impact on polarization purity.

Finally, we note that for the warm radiometer optics, even as deployed and including all misalignments (green curve), the leakage is still far below $r=0.001$ after scaling to 95\,GHz. The optics appear to have sufficient polarization purity as they are currently deployed, and this will further be improved by fixing the most important misalignments. Even though the leakage is still small, it is worth noting that it is beneficial to fix the radome sag and any other misalignments that break circular symmetry. The optics are significantly faster to simulate when they're circularly symmetric and other operations (map making, correcting \TtoP\ leakage) can be less expensive for circularly symmetric optics. We will likely try to fix all of these misalignments for the cryogenic version of CGEM.

\subsection{Solar Beam Measurements}\label{sec:solar}
\subsubsection{Model for Solar Data}
Here we compare the beam model described in the previous section to a measurement of the beam using an hour-long Solar observation from July 18th, 2025. The Sun has an angular size that is comparable to our beam FHWM, and so it must be modelled carefully.

We make a forward model of the time-ordered data from the beam model and estimates of the gain and noise bias. Equation \ref{eq:antenna_voltage} describes how the optics couple sky signals into the circular waveguide after the feedhorn. Our waveguide network (including the OMT) and analog radiometer then effectively apply Jones matrices to these signals before they are correlated. We accordingly model the time-ordered data as
\begin{align}
	C_{11} &= g_1(t)g_1^*(t) (J_{1r}J^*_{1r}W_{rr}(t) + J_{1r}J^*_{1l}W_{rl}(t) + J_{1l}J^*_{1r}W_{lr}(t) + J_{1l}J^*_{1l}W_{ll}(t) + N_{11})+ n_{11}\nonumber\\
	C_{12} &= |g_1(t)g_2(t)^*|e^{i\delta(t)}(J_{1r}J^*_{2r}W_{rr}(t) + J_{1r}J^*_{2l}W_{rl}(t) + J_{1l}J^*_{2r}W_{lr}(t) + J_{1l}J^*_{2l}W_{ll}(t) + N_{12})+ n_{12}\label{eq:high_level_TOD_model}\\
	C_{22} &= g_2(t)g_2^*(t) (J_{2r}J^*_{2r}W_{rr}(t) + J_{2r}J^*_{2l}W_{rl}(t) + J_{2l}J^*_{2r}W_{lr}(t) + J_{2l}J^*_{2l}W_{ll}(t) + N_{22}) + n_{22}.\nonumber
\end{align}
The $J_{ip}$ are the Jones matrix elements for the waveguide network, where $J_{ip}$ couples the antenna signal $W_p$ into channel $i$. As written above, $C_{11}$ predominantly measures right-hand circular polarization, while $C_{22}$ predominantly measures left-hand circular polarization. $g_1(t)$ and $g_2(t)$ are the complex gains of each radiometer channel and are the Jones matrix elements of the analog radiometer, which is assumed to be diagonal. $\delta(t)$ is the phase difference between the channels. $N_{ij}$ are noise bias terms and $n_{ij}$ are random noise terms. The noise bias terms could be time-dependent in general, but we assume they are constant here. $N_{12}$ is small ($|N_{12}| \ll N_{11},N_{22}$) but measured to be non-zero and frequency-dependent in the CGEM data. We attribute this to a small amount of leakage between the two mixers in the radiometer, and we are reconfiguring the cold radiometer to reduce this substantially in the future. Note that in the following, we only model the magnitude of $C_{12}$, as we have not yet calibrated the $\delta(t)$ at high signal to noise.

We use lab-bench measurements of the $J_{ip}$ in our model, we measure $|g_1(t)g^*_1(t)|$ and $|g_2(t)g^*_2(t)|$ from calibrator sources, and we estimate the $N_{ij}$ noise bias terms by median filtering the Solar data over time spans in the data where the Sun is not in the beam. To model the $W_{pq}$ terms for the Sun, we start by calculating the beam with TICRA Tools (including the misalignments discussed in Section~\ref{sec:misalignments_and_model}). We assume the Sun is unpolarized ($Q = U = V = 0$) and is a disk of constant surface brightness. The data we examine here from July 18th, 2025 had only moderate solar flux density ($\sim$160\,sfu), indicating fairly low activity (and hence low polarization and few Sun spots). We take the Sun to have angular diameter in our band equal to the optical angular diameter \cite{solar_radius_and_brightness_temp} on the day of the observations. We also assume that the Sun dominates all other sky and ground emission in the time-ordered data. This is good approximation, as the Sun has an antenna temperature of thousands of Kelvin, which is significantly larger than the sky brightness (order $10$\,mK) and ground pickup (which varies by at most $\sim$100\,mK in azimuth, due to an open panel in the CGEM ground screen, which has since been closed). We can hence model the beam terms as
\begin{align}
	W_{pq}^{\text{disk}}(\nu, t) &= \int d^2\n\big[B_{pq}^{T} (\nu, \n) \, T_\text{sun}(\nu)\Theta(\n'(t) - \n)\Big],\label{eq:sun_W_pq_model}
\end{align}
where $T_\text{sun}(\nu)$ is the (assumed spatially constant) surface brightness of the Sun, $\n'(t)$ is the location of the Sun in the telescope frame, and $\Theta(\n)$ is a step function to define the solar disk:
\begin{equation}
	\Theta(\n) = \begin{cases}
				1 & \theta \leq \theta_\text{sun}\\
				0 & \text{otherwise}.
			   \end{cases}
\end{equation}
Upon inspection, it is clear that Equation~\ref{eq:sun_W_pq_model} is a convolution on the sphere with the circularly symmetric function $\Theta(\n)$. To forward model the data, we can therefore precompute the beam transfer function convolved with a $0.5^\circ$ disk, and evaluate it at the location of the Sun in the telescope frame at every time step. We do the convolution in spherical harmonic space for computational efficiency. We then bin the time-ordered data and model into maps, and fit the model map to the data with $T_\text{sun}(\nu)$ as the only parameter. This merely scales the a priori model map by a single number.

\subsubsection{Solar Data and Model Comparison}
\begin{figure} [ht]
   \begin{center}
   \begin{tabular}{c} 
   \includegraphics[height=8.3cm]{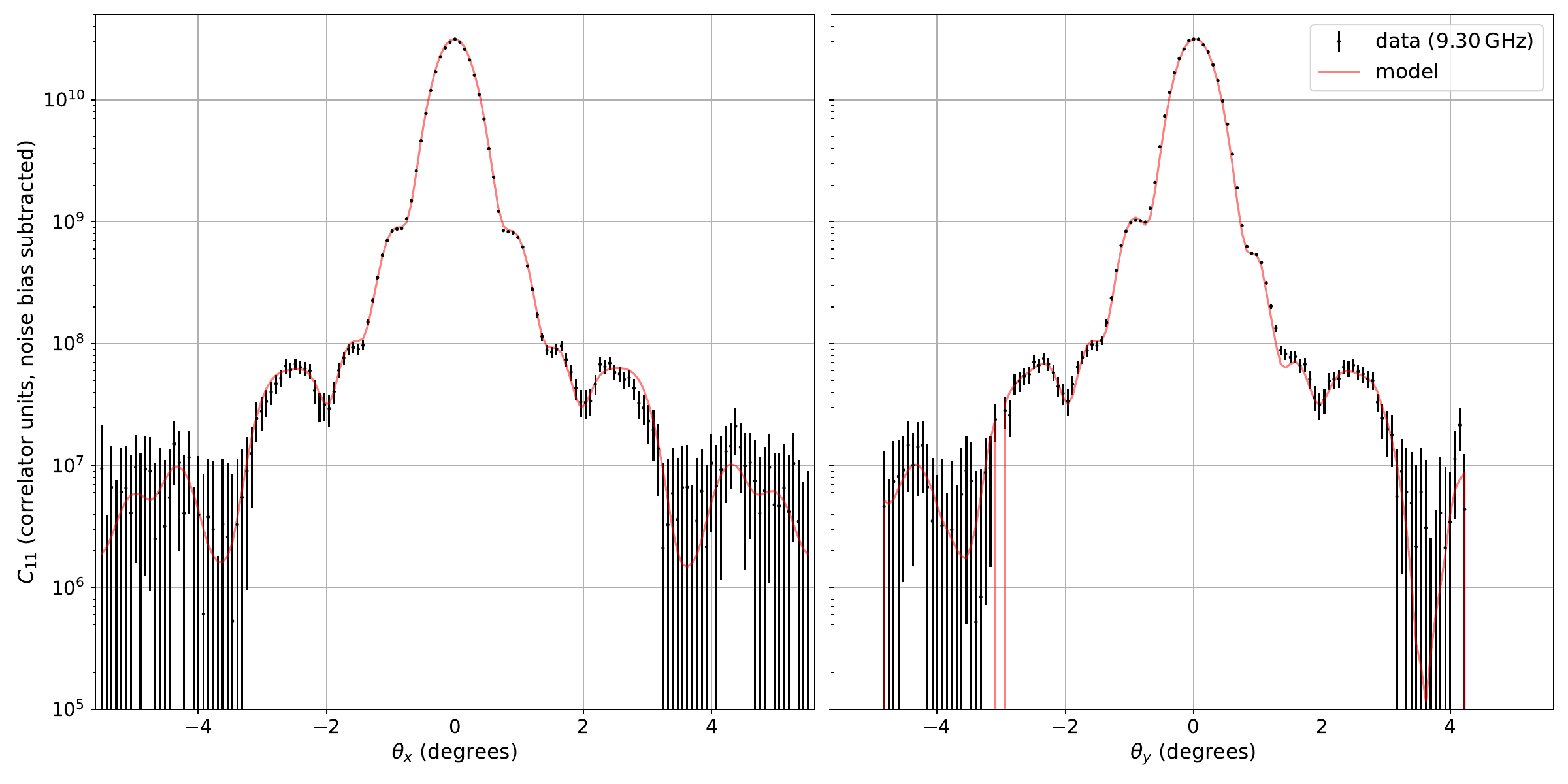}
   \end{tabular}
   \end{center}
   \caption[example] 
   {Slices through data and model maps of the $C_{11}$ autocorrelation for the Solar observation on July 18th, 2025. The data are plotted in uncalibrated correlator units, with the noise bias $N_{11}$ subtracted. The left panel shows a slice through the beam parallel to the scan direction, while the right panel shows a slice perpendicular to the scan direction (in the elevation direction). Negative $\theta_y$ values are at lower elevation than the boresight. Due to the finite angular extent of the Sun, this is a low-resolution probe of the beam, in which we see an excellent agreement between the data and model. The model shown here accounts for the measured misalignments discussed in Section~\ref{sec:misalignments_and_model}, but is otherwise an a priori beam calculation. Note that there is an unobserved pixel in the elevation slice (right panel), which is why the data and model drop to 0 near $\theta_y = -3^\circ$.\label{fig:solar_beam_slices}}
\end{figure}
Figure \ref{fig:solar_beam_slices} shows azimuth (left panel) and elevation (right panel) cuts through the data and model maps for the $C_{11}$ autocorrelation at 9.3\,GHz. The mean background level in the maps (which is non-zero because of the $N_{11}$ noise bias term) is subtracted prior to plotting. This plot compares the measured and model power beam at low angular resolution (due to the finite angular extent of the Sun). The corresponding plot made for $C_{22}$ is identical in nature. We see a strong agreement between the data and model over nearly 4 orders of magnitude. Making the same maps and plots at other frequencies yields a similarly striking agreement. We can see beam asymmetries in the elevation slice, which are due to the optical misalignments in Section~\ref{sec:misalignments_and_model}. The asymmetries are well-captured by the model. Slices through other directions in the map show a similar agreement. It can be seen in the elevation slice that the measured beam is slightly wider than in the model. We believe this is due to modelling the radome sag under gravity as a simple $0.4^\circ$ tilt of the radome about its base: the true radome deformation is likely more complicated. In the near future, we expect to either take more measurements to improve the model or to correct the radome sag (e.g. with shims or making a stiffer Astroquartz support).

\begin{figure} [ht]
   \begin{center}
   \begin{tabular}{c} 
   \includegraphics[width=10.5cm]{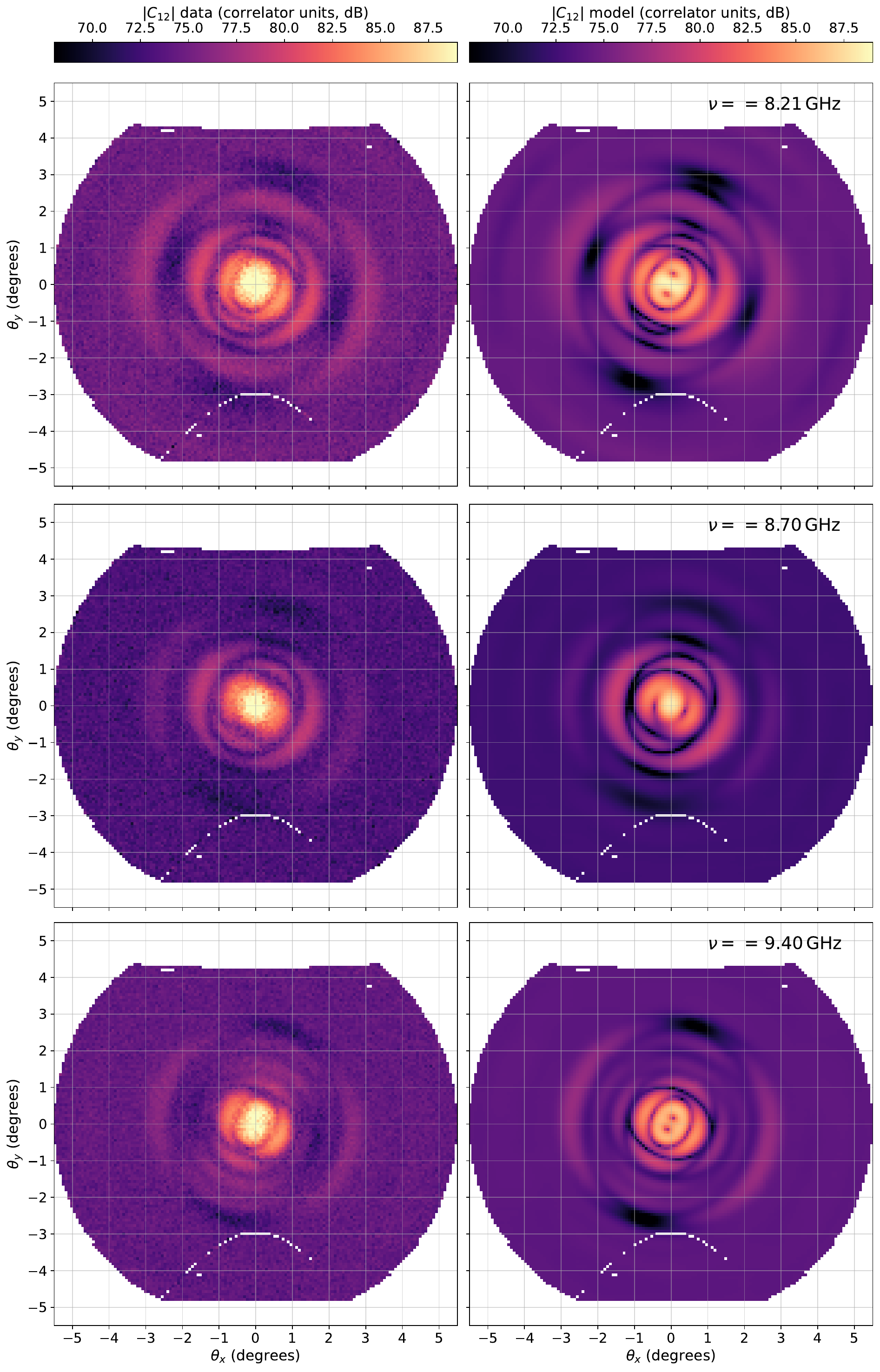}
   \end{tabular}
   \end{center}
   \caption[example] 
   {$|C_{12}|$ maps for the Solar data (left column) and model (right column) at three representative frequencies in the band (frequencies are labelled in the top right corner of each plot in the right column). The $|C_{12}|$ data probe $T\rightarrow P$ leakage, and the model agrees very closely at all frequencies. This suggests that we have an excellent model for our most important beam systematic. We remind the reader that the model shown here accounts for the measured misalignments discussed in Section~\ref{sec:misalignments_and_model}, but is otherwise an a priori beam calculation. Note that these signals are $\sim$18\,dB down from the peak signal in the autocorrelations\label{fig:delta_P_maps_multifreq}.}
\end{figure}
Figure~\ref{fig:delta_P_maps_multifreq} shows data (left column) and model (right column) maps of $|C_{12}|$ at three frequencies: 8.21\,GHz (top row), 8.70\,GHz (middle row), and 9.4\,GHz (bottom row). Because of the low solar activity on the day that these data were taken, we expect the Sun's emission to be unpolarized. These maps are therefore low-angular-resolution maps of the \TtoP\ leakage beam. The data and model are in striking agreement at all frequencies, and nearly all qualitative features in the data are also present in the model. The model clearly captures the frequency evolution of the \TtoP\ leakage. This suggests that we have an excellent model for the \TtoP\ leakage of the deployed optics. We may be able to use this model to remove polarization leakage from our data in future. We note that the model still slightly underestimates the response in $|C_{12}|$ near boresight. This could also arise if the Sun had a small amount of $V$ polarization from (small amounts of) Solar activity. We expect to further investigate and resolve this discrepancy with more measurements and observations in the coming months.

\subsection{Probing the Beam with Satellites}
The CGEM radio-frequency interference (RFI) environment is pristine, with less than 1\% of our data being flagged by kurtosis-based RFI excision algorithms. The RFI that we do see predominantly comes from Earth-observation satellites downlinking their data in our band below $\sim$8.4\,GHz. Fortunately, these satellites are incredibly useful beam measurement tools: we see one every $\sim$10\,minutes, they appear as constant emitters in our data (sampled at 1\,ms cadence), are often highly circularly polarized, and emit strong enough signals to probe our beam down to 40\,dB below the peak. Their bright emission in combination with our high scan rate also means that the beam is probed to the -40\,dB level on $< 1$\,s time scales, over which instrumental properties (gains, etc.) are constant, making the modelling simpler. The downside of these sources as beam probes is that we do not know the polarization properties (Stokes parameters) of their emission. We may be able to get around this in future analyses if satellite companies are willing to share their down-link antenna beam profiles/transmission properties, or perhaps by jointly analyzing an ensemble of satellite beam crossings. For now, we fit for the satellite Stokes parameters in our model, as described below.

\subsubsection{Satellite Beam Validation Procedure}\label{sec:sat_beam_validation_procedure}
We use the same model as in Equation~\ref{eq:high_level_TOD_model} to model satellite observations, but update the $W_{pq}$ terms (recalling that satellites are point sources when observed with a 0.5$^\circ$ beam):
 \begin{align}
	W^{\text{sat}}_{pq}(t) &= \frac{c^2}{2k_B\nu^2} \int d^2\n\big[B_{pq}^{T}(\n)\delta(\n'(t) - \n) S^T_{\text{sat}} + B_{pq}^{Q}(\n)  \delta(\n'(t) - \n) S^Q_{\text{sat}} \nonumber\\ &\quad\quad\quad\quad + B_{pq}^{U}(\n) \delta(\n'(t) - \n) S^U_{\text{sat}} + B_{pq}^{V}(\n)  \delta(\n'(t) - \n) S^V_{\text{sat}}\big]\nonumber\\
	&= \frac{c^2}{2k_B\nu^2} \big[B_{pq}^{T}(\n'(t))S^T_{\text{sat}} + B_{pq}^{Q}(\n'(t)) S^Q_{\text{sat}} + B_{pq}^{U}(\n'(t)) S^U_{\text{sat}} + B_{pq}^{V}(\n'(t)) S^V_{\text{sat}}\big],
\end{align}
where $S^X_\text{sat}$ is the flux density from the satellite (e.g. in W\,m$^{-2}$\,Hz$^{-1}$) in Stokes parameter $X$ and the factor in front of the integral gives the total expression units of Kelvin. We can clearly see that satellites trace the beam transfer functions at full angular resolution, unlike the Sun. To evaluate $W_{pq}$, we calculate the beam with TICRA Tools (including the misalignments discussed in Section~\ref{sec:misalignments_and_model}), compute the beam transfer functions ($B_{pq}^X(\n)$), and evaluate them at the location of the satellite (relative to the boresight) at each time step. We once again only model $C_{11}$, $|C_{12}|$, and $C_{22}$.

Since the $S_\text{sat}^X$ are unknown, we fit the model to the $C_{11}$, $|C_{12}|$, and $C_{22}$ data simultaneously with the $S_\text{sat}^X$ as parameters. For this reason, this analysis can only be used as a \textit{validation} of the existing beam model, and not as a beam \textit{measurement}. This is still a strong validation: we are simultaneously fitting $C_{11}$, $|C_{12}|$, and $C_{22}$, each of which have complicated time-dependence, with only 4 parameters that add fixed functions ($B_{pq}^X$) in a linear combination for each correlation product. These functions cannot produce arbitrary sidelobe structure in $C_{11}$, $|C_{12}|$, and $C_{22}$ simultaneously and it is hence unlikely that the Stokes parameter fitting creates an ``artificial" agreement between the data and model. If the model simultaneously agrees well with the $C_{11}$, $|C_{12}|$, and $C_{22}$ satellite data after fitting for the $S_\text{sat}^X$, this tells us that the beam model is generally consistent with the data, that the agreement in the plots is a ``best case" agreement, and that other (generally small) discrepancies could exist and may be hidden by the Stokes parameter fitting. If the beam model is \emph{not} able to capture the data, this tells us that the beam model has deficiencies that need to be further investigated. The agreement between the data and model has been excellent for the set of $\sim$30 satellites that we have examined.

To validate the beam model with satellites, the procedure is:
\begin{itemize}
	\item Examine waterfall plots of CGEM data and identify satellites by looking for bright RFI below $\sim$8.4\,GHz.
    \item Find the peak time and frequency of a satellite signal of interest, or if the satellite is broadband, find the frequency range over which there is significant emission.
	\item Based on the time and boresight location on the sky, query Satchecker\cite{satchecker} for satellites within a 1$^\circ$ field-of-view (FOV) around the boresight. We have only ever observed this query to return exactly one satellite. Record the North American Aerospace Defence Command (NORAD) ID of the satellite.
	\item With the satellite NORAD ID, query Satchecker to find the nearest-issued two line orbital elements (TLEs) for the satellite.
	\item Use the TLEs to compute the altitude and azimuth coordinates of the satellite as seen from the CGEM site, and transform them into the beam coordinates. This gives us the track of the satellite through the beam and we can now evaluate the model $W_{pq}^\text{sat}$ terms and hence $C_{11}$, $|C_{12}|$, and $C_{22}$.
	\item Fit the model $C_{11}$, $|C_{12}|$, and $C_{22}$ to the data. The fit parameters are the $S_\text{sat}^X$ for each stokes parameter (4 parameters) and a shift of the model in time to pick up (small) TLE errors and (small) timing/pointing errors from CGEM.
	\item Compare the model fit and data to validate the beam model.
\end{itemize}
Note that much of this procedure will be automated in future.

\subsubsection{Satellite Data and Beam Model Comparison}

\begin{figure} [ht]
   \begin{center}
   \begin{tabular}{c} 
   \includegraphics[height=12.75cm]{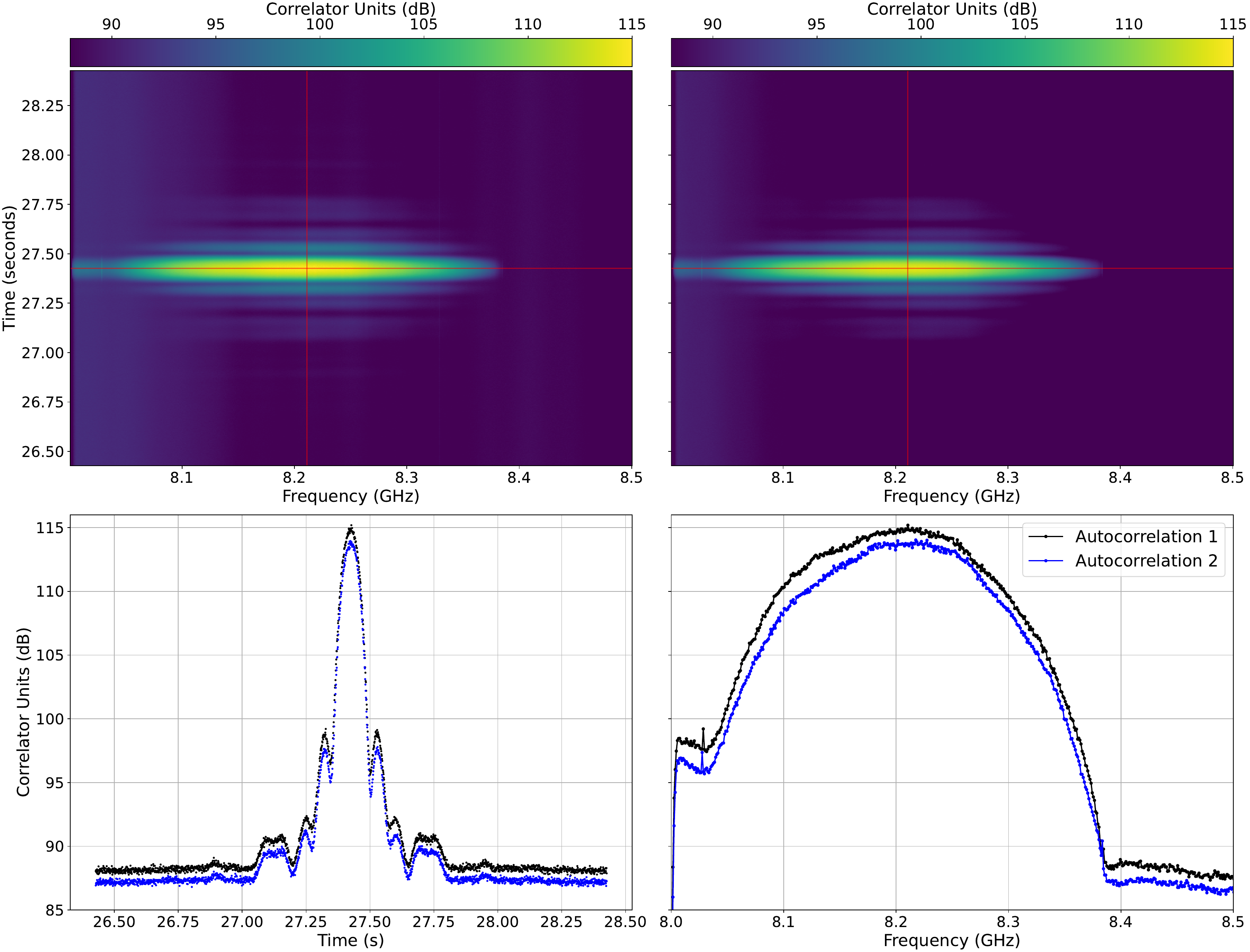}
   \end{tabular}
   \end{center}
   \caption[example] 
   {A snapshot of CGEM data at the time that we saw the SWOT satellite on the night of October 6th, 2025. The top left and right plots show waterfalls for $C_{11}$ and $C_{22}$, respectively. The horizontal red line indicates the time of the peak signal measured by CGEM. The red vertical line indicates the peak frequency of the emission (8.2105\,GHz, which is the frequency that we have used to validate the beam model in Figure~\ref{fig:SWOT_fits_peak_freq}. The bottom left plot shows a time slice at the peak frequency, $8.2105$\,GHz and the bottom right plot shows a frequency slice at the time of the peak measured signal. There is appreciable signal over nearly 400\,MHz of bandwidth, and both transmitters (RCP and LCP) appear to be active.\label{fig:SWOT_data_overview}}
\end{figure}
Here we showcase just one validation of the beam model from a single satellite observation: an observation of the NASA Surface Water and Ocean Topography (SWOT) satellite \cite{swot} on October 6th, 2025. The satellite has two circular polarized downlink antennas: one transmits in RCP and the other in LCP. The beams of these downlinks are ``isoflux" and have the angular size of the Earth from the viewpoint of the satellite. Figure~\ref{fig:SWOT_data_overview} shows an overview of how the satellite looked in the data. The top row of plots shows waterfalls of the data in each autocorrelation. The bottom left plot shows a time slice at the peak frequency of the emission (8.2105\,GHz, shown as the red horizontal line in the top row of Figure~\ref{fig:SWOT_data_overview}) and the bottom right plot shows a frequency slice at the time of the peak measured signal (shown as the central red vertical line in the top row of Figure \ref{fig:SWOT_data_overview}). We received approximately the same signal level in both channels (the differences in Figure~\ref{fig:SWOT_data_overview} are roughly consistent with the known difference in noise bias levels between the two channels), so it is likely that both of the RCP and LCP downlinks were active at the time we saw the satellite. We can also see that the satellite is very broad band: it emits with appreciable signal over nearly 400\,MHz of bandwidth. The peak signal is $\sim$30\,dB above the noise floor at the peak frequency. Note that we saw no evidence of non-linearity or overflows in the data for this satellite observation.

\begin{figure} [ht]
   \begin{center}
   \begin{tabular}{c} 
   \includegraphics[width=16.7cm]{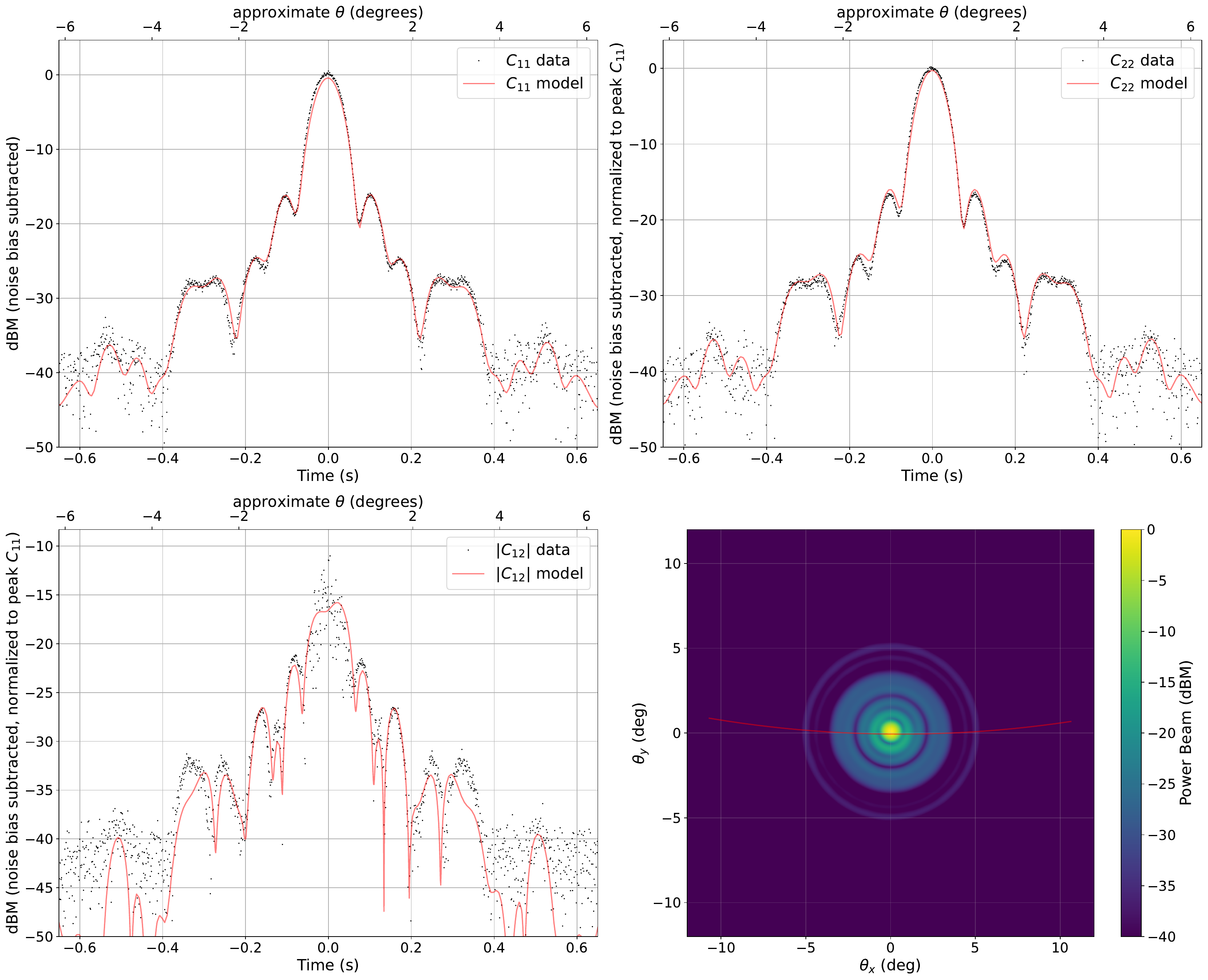}
   \end{tabular}
   \end{center}
   \caption[example] 
   {SWOT satellite data/model comparison plots. The model is derived from an electromagnetic calculation of the beam that includes the misalignments discussed in Section~\ref{sec:misalignments_and_model}. The top plot shows an image of the power beam with the SWOT track through the beam overlaid as the red curve. The 2nd row shows $C_{11}$ (left) and $C_{22}$ (right) plotted in correlator units. The 3rd row shows the same data as the first row, but with the noise bias subtracted and peak normalized to better highlight beam structure. The bottom row shows $|C_{12}|$ in correlator units (left plot) and with the noise bias subtracted and peak normalized to $C_{11}$ (right plot). All time-series plots have a second axis indicating the approximate distance of the satellite from boresight. All plots show an excellent agreement between the data and model, with small discrepancies discussed in the text.\label{fig:SWOT_fits_peak_freq}}
\end{figure}

Figure~\ref{fig:SWOT_fits_peak_freq} shows a comparison between the data and model at the peak frequency, $8.2105$\,GHz. The bottom-right panel shows a map of the model power beam, with a red curve overlaid that represents the track that the satellite took through the beam. The track was derived from the TLE method described in Section~\ref{sec:sat_beam_validation_procedure}. The satellite passed within 0.07$^\circ$ of boresight. The other three plots show the data and model for $C_{11}$ (top left), $C_{22}$ (top right), and $|C_{12}|$ (bottom right), after subtracting an estimate of the noise bias and peak-normalizing the data to the peak signal in $C_{11}$. All data/models are plotted with time on the $x$ axis, but there is a 2nd $x$ axis at the top of each plot that indicates the approximate angular distance of the satellite from boresight (this is approximate because it assumes the satellite was moving uniformly in $\theta$ in time).

The $C_{11}$ and $C_{22}$ plots in Figure~\ref{fig:SWOT_fits_peak_freq} show excellent agreement between the data and model. The main beam peak in the model is slightly lower than in the data, which might be attributable to TLE errors or an overestimation of the sidelobe level in the model. The latter is possible, as the beam model shown here was primarily informed by low-resolution observations of the Sun. We note that this satellite probes structure that is $40$\,dB down from the peak in the autocorrelations. The Stokes parameters from the fit suggest that the satellite was $\sim$$2\%$ circularly polarized and $\sim3\%$ linearly polarized. The bottom-left panel therefore primarily compares \TtoP\ leakage in the data and model, and we can see a strong agreement between the two. Nearly all sidelobes present in the data are also present in the model, however their relative heights are off in several places by $\sim$1\,dB. We also note that it appears that the on-axis response in $|C_{12}|$ (presumably leakage) may be underestimated in the model (this can also be seen in $|C_{12}|$ data/model maps of the Sun in Figure~\ref{fig:delta_P_maps_multifreq}). However, considering that this beam model was derived from measurements of the Sun in $C_{11}$ and $C_{22}$ ($|C_{12}|$ was only used as a qualitative validation) and that we are probing low-level leakage 40\,dB down from the peak in $C_{11}$, this agreement is quite encouraging. We expect to resolve many of these discrepancies with future beam characterization work. Finally, note that we are probing structure in \TtoP\ leakage out to $\sim$4$^\circ$ off boresight, corresponding to angular scales out to $\ell \sim 45$ (even larger than the ``recombination bump" angular scales near $\ell \sim 80$ that are relevant to the \bmode\ search). The strong agreement between the data and model in $C_{11}$, $|C_{12}|$, and $C_{22}$ is similar at other frequencies. This further suggests that we have an excellent model for the deployed optics.

\section{Conclusions}\label{sec:conclusions}
In this paper, we have described the CGEM optical design and on-sky characterization effort in detail. We described the hat feed optical design employed by CGEM, which is the first of its kind in cosmology and radio astronomy and exhibits excellent polarization purity. We then described a novel optimization framework that was used to shape the optics to minimize simulations of our most important beam systematic, $T\rightarrow P$ leakage, in angular power spectrum space. This framework could be used by other instruments to minimize beam systematics in the optical design phase. The resulting optimized hat feed design has simulated $T\rightarrow P$ leakage angular power spectra that are orders of magnitude below $C_{\ell}^{BB}$ at $r = 10^{-3}$ when scaled to the CMB observing window near 95\,GHz. We then described the mechanical design of the CGEM optics, including a novel BOR secondary mirror support made from low loss, low dielectric constant, Astroquartz composite. We briefly described the procedure that we used to measure several small, correctable misalignments in the optics. These misalignments are simulated to have negligible effect on polarization purity. We then showcased the fantastic agreement between our beam model (derived from electromagnetic calculations of the optics) and measurements of the beam from observations of the Sun and Earth-observation satellites. These sources probe the beam several orders of magnitude down from the peak and out to $\sim$5$^\circ$ off the boresight, and the model is an excellent match to the data throughout. These early observations show immense promise for future science with CGEM.

In the coming months, we plan to fix several of the measured optical misalignments: the primary mirror distortion, the feed column $\Delta z$ defocus, and the feed column tilt. We also plan to explore fixing the sag of the secondary mirror support, which may require manufacturing a new part. We expect these adjustments to bring the performance even closer to the design performance. We aim to press forward with the on-sky characterization effort by mapping traditional astrophysical point sources (e.g. Tau A and Cas A), and continuing to refine beam measurements derived from observations of the Sun and Moon (the latter of which was not described here). We also hope to improve the satellite beam measurement procedure, perhaps by learning the emission properties of a handful of satellites, doing a joint analysis of a large ensemble of satellite observations, or perhaps by taking observations with an accompanying antenna with a known beam pattern (e.g. Ref.~\citenum{sat_beam_meas}). We also aim to explore techniques for using our beam models to correct $T\rightarrow P$ leakage in the time-ordered data to further enhance the polarization purity of our maps. Lastly, we plan to incorporate what we have learned in the optical design and on-sky characterization effort into the design of the cryogenic system optics, for which we expect to obtain even better polarization purity.

\appendix    

\acknowledgments 

CGEM is located on the traditional, ancestral and unceded territory of the Okanagan Syilx people, within the grounds of the Dominion Radio Astrophysical Observatory (DRAO). We benefit enormously from working on these lands. We thank the Dominion Radio Astrophysical Observatory, operated by the National Research Council Canada, for gracious hospitality and expertise. We thank the UBC machine shop, Spin Industries Ltd., Sightline Engineering, and Waycon Manufacturing Ltd. for bringing crucial components of CGEM to life.

CGEM was funded by a grant from the Canada Foundation for Innovation (CFI) 2018 Leading Edge Fund (Project 36483) and  by contributions from the province of British Columbia and the Natural Science and Engineering Research Council (NSERC) through Discovery Grants. This research was enabled in part by support provided by CalculQuebec \href{(https://www.calculquebec.ca/en/)}{https://www.calculquebec.ca/en/} and the Digital Research Alliance of Canada \href{(alliancecan.ca)}{alliancecan.ca}. We gratefully acknowledge the Collaboration for Astronomy Signal Processing and Electronics Research (CASPER) for providing open-source hardware and software development tools
that were essential to the signal processing infrastructure used in this research. J.M. acknowledges the support of the Natural Sciences and Engineering Research Council of Canada (NSERC) [funding reference number 569654]. P.V.G. received the support of a fellowship from “la Caixa” Foundation (ID 100010434,  code  B005800), a Rafel del Pino Excellence Scholarship and a UBC 4YF Doctoral Fellowship. P.Z. received the support of a UBC 4YF Doctoral Fellowship.

\bibliography{report} 
\bibliographystyle{spiebib} 

\end{document}